\documentclass{aa}  

\usepackage{graphicx}
\usepackage{txfonts}
\usepackage{lipsum}
\usepackage{subcaption}         % necessary for continued figures, example in section 3
\usepackage{lscape}             % to rotate a single page table, example in appendix.
\usepackage{placeins}           % useful with \FloatBarrier, to keep 
\usepackage[colorlinks=true,citecolor=blue,linkcolor=blue,urlcolor=blue]{hyperref}
\usepackage[all]{hypcap}
\usepackage{xcolor}
\usepackage[utf8]{inputenc}
\usepackage{lineno}
\usepackage{color}
\usepackage{verbatim} 

\newcommand{\Ha}{H$\alpha$}
\newcommand{\Hb}{H$\beta$}
\newcommand{\OH}{$12+\log$(O/H)}

\begin{document}

%%%%%%%%%%%%%%%%%%%%%%%%%%%%%%%%%%%%%%%%
% if you use custom commands in your title,
% ensure to check your title when submitting!
%%%%%%%%%%%%%%%%%%%%%%%%%%%%%%%%%%%%%%%%
   \title{ALMA Chemical Evolution (ACE) survey: an overview}

   \subtitle{Extending dust and gas inference to low metallicities at cosmic noon}

%%%%%%%%%%%%%%%%%%%%%%%%%%%%%%%%%%%%%%%%
% Please separate each author with the \and command
%
% Please do not include ORCIDs next to author names.
% Only ORCIDs authenticated by individual authors in EDPS
% editorial system will be taken into account.
% ORCIDs included here will be removed.
%%%%%%%%%%%%%%%%%%%%%%%%%%%%%%%%%%%%%%%%

   \author{Irene Shivaei\inst{1}\fnmsep\thanks{Corresponding author: ishivaei@cab.inta-csic.es}
        \and 
        Gerg\H{o} Popping\inst{2}
        \and
        Leindert A. Boogaard\inst{3}
        \and
        Alexandra Pope\inst{4}
        \and
        Ivanna Langan\inst{1}
        \and
        Roxana Popescu\inst{4}
        \and
        Nikki N. Geesink\inst{2}
        \and 
        Manuel Solimano\inst{1}
        \and
        %alphabetical order
        Caitlin M. Casey\inst{5,6}
        \and
        Maximilien Franco\inst{7}
        \and
        Melanie Kaasinen\inst{8}
        \and
        Bahram Mobasher\inst{9}
        \and
        Desika Narayanan\inst{10,6}
        \and
        Pablo G. P{\'e}rez-Gonz{\'a}lez\inst{1}
        \and
        Naveen Reddy\inst{9}
        \and
        Ryan Sanders\inst{11}
        }

   \institute{Centro de Astrobiolog\'ia (CAB), CSIC-INTA, Ctra. de Ajalvir km 4, Torrej\'on de Ardoz, E-28850, Madrid, Spain
   \and
   European Southern Observatory, Karl-Schwarzschild-Str. 2, D-85748, Garching, Germany
    \and
    Leiden Observatory, Leiden University, PO Box 9513, NL-2300 RA Leiden, The Netherlands  
    \and
    Department of Astronomy, University of Massachusetts, Amherst, MA 01003, USA
    \and
    Department of Physics, University of California, Santa Barbara, Santa Barbara, CA 93106, USA
    \and
    Cosmic Dawn Center at the Niels Bohr Institute, University of Copenhagen and DTU-Space, Technical University of Denmark
    \and
    Universit{\'e} Paris-Saclay, Universit{\'e} Paris Cit{\'e}, CEA, CNRS, AIM, 91191, Gif-sur-Yvette, France
    \and
    Research School of Astronomy and Astrophysics, Australian National University, Canberra, ACT 2611, Australia
    \and
    Department of Physics and Astronomy, University of California, Riverside, 900 University Avenue, Riverside, CA 92521, USA
    \and
    Department of Astronomy, University of Florida, 211 Bryant Space Sciences Center, Gainesville, FL 32611 USA
    \and
    Department of Physics and Astronomy, University of Kentucky, 505 Rose Street, Lexington, KY 40506, USA
              }

   \date{Submitted to A\&A}

% \abstract{}{}{}{}{}
% 5 {} token are mandatory
 
  \abstract
  {The baryon cycle is central to galaxy evolution, governing the exchange of gas, metals, and dust between galaxies and their surrounding environments. Yet, simultaneous constraints on these constituents remain scarce beyond $z\sim1$. The ALMA Chemical Evolution (ACE) survey is a Cycle 11 ALMA Large Program designed to address this by studying the molecular gas, dust, and metal content of sub-solar metallicity at cosmic noon. ACE consists of CO(3-2) and dust continuum observations of 25 galaxies at $z=2.0-2.5$, drawn from the MOSDEF survey, with robust gas-phase metallicity measurements spanning $\sim$35\% to 8\% of solar metallicity. The survey approximately doubles the number of unlensed main-sequence galaxies at z>1 with CO, dust-continuum and metallicity measurements and extends such studies to almost an order of magnitude lower stellar masses and metallicities than previous surveys. The ACE observations yield 17 CO detections and 17 Band 7 continuum (rest-frame $\sim 270\,\mu$m) detections. CO detectability correlates most strongly with metallicity and stellar mass, while dust continuum detectability is more closely linked to star formation rate and infrared luminosity. Using the ACE measurements, we derive a new empirical scaling relation linking CO(3-2) luminosity to stellar mass, star formation rate, and metallicity, providing a practical benchmark for estimating molecular gas content in low-mass, low-metallicity galaxies. The survey reveals several particularly intriguing systems, including some of the lowest-metallicity CO and dust detections currently known at $z>1$, galaxies with extreme gas and dust fractions, and systems exhibiting offsets between stellar, dust, and molecular gas emission. ACE provides the first comprehensive view of the interplay between molecular gas, dust, metals, and star formation in sub-solar metallicity galaxies at cosmic noon, enabling direct tests of models for baryon cycling, chemical enrichment, and dust evolution during the peak epoch of galaxy assembly.}  

   \maketitle
   \nolinenumbers

%%%%%%%%%%%%%%%%%%%%%%%%%%%%%%%%%%%%%%%%%%%%%%%%%%%%%%%%%%%%%%
\section{Introduction}

The cycling of baryons in and out of galaxies regulates their formation and evolution across cosmic time. In this baryon cycle, gas accretion from the cosmic web fuels star formation, while stellar and active galactic nucleus feedback drives outflows that redistribute gas and metals into the circumgalactic medium before eventual re-accretion \citep{Bouche2010,dave12, Lilly2013}. This interplay sets the gas content, star-formation rate (SFR), and chemical enrichment of galaxies and is therefore central to models of galaxy evolution \citep{somerville15}. The importance of this cycle is particularly evident during ``cosmic noon'' ($z \sim 1$--$3$), the epoch of peak cosmic star-formation activity and molecular gas density \citep{madau14,peroux20}. Galaxies at this epoch exhibit elevated gas fractions, enhanced star formation efficiencies, and different interstellar medium (ISM) conditions compared to the local Universe \citep{steidel14,tacconi18,Aravena2019,sanders21}, underscoring the need for direct observations of representative galaxy populations at high redshift.

A complete empirical description of the baryon cycle requires simultaneous constraints on all major baryonic components: molecular gas as the fuel for star formation, SFRs tracing ongoing stellar buildup, stars as the integrated outcome of the star-formation history, and metals and dust recording the cumulative chemical enrichment history. However, observational studies at $z \sim 2$ have traditionally focused on individual components rather than a unified view. Large surveys have established molecular gas scaling relations using CO and dust continuum tracers \citep{tacconi13, scoville16, Liu2019, Aravena2019, aravena20}, and dust scaling relations from Herschel data \citep[e.g.,][]{reddy12a,schreiber15,kirkpatrick17}. Yet, these studies often lack robust gas-phase metallicities for the same galaxies. 
A major reason for this difficulty is that it is comparatively much easier to obtain rest-UV and optical observations of every very faint, low mass, and low-metallicity galaxies at high redshift (giving access to diagnostics of the ISM and stellar populations), while dust and molecular observations of the same galaxies are much harder to obtain due to both sensitivity limits and, for some facilities, confusion, and also intrinsically fainter cold gas and dust emission in low-metallicity galaxies owing to evolving $\alpha_{\rm CO}$ and dust-to-gas ratio with metallicity \citep{remyruyer14,boogaard21}.
As a consequence, samples with a complete baryonic inventory---gas, stars, SFR, metals, and dust--- are extremely rare at cosmic noon, particularly in the regime of typical, sub-solar metallicity galaxies.

Scaling relations linking galaxy properties provide a powerful framework to interpret the baryon cycle. The mass--metallicity relation (MZR) and its extension, the fundamental metallicity relation (FMR), connect stellar mass, SFR, and gas-phase metallicity, reflecting the equilibrium between gas inflow, star formation, and outflows \citep{Mannucci2010, sanders21}. A more physically motivated extension is the gas fundamental metallicity relation (GFMR), which incorporates the molecular gas reservoir and links $M_\star$, SFR, metallicity, and molecular gas mass ($M_{\mathrm{mol}}$), thereby directly probing the interplay between gas flows and chemical enrichment \citep{Bothwell2016}. In this framework, inflows of metal-poor gas increase gas fractions and SFRs while diluting metallicity, whereas feedback-driven outflows remove enriched material, regulating the chemical evolution of galaxies \citep{Dave2017}. While scaling relations coupling gas mass and metallicity have been studied at low redshift \citep{Bothwell2016}, extending them to lower stellar masses and metallicities at $z \sim 2$ remains a major observational challenge (c.f., e.g., \citealt{Saintonge2013, Catan2024}).

Dust provides an additional and critical probe of the baryon cycle by linking metals in the gas phase to those locked into solid grains, while regulating the thermodynamics and chemistry of the ISM. The derivation of dust masses relies on modeling the far-infrared (FIR) and submillimeter spectral energy distribution (SED), where uncertainties in the dust emissivity index ($\beta$) and mass-weighted dust temperature ($T_{\rm dust}$) remain a dominant source of systematic error. In the local Universe, both the shape of the dust SED and the effective values of $\beta$ and $T_{\rm dust}$ are observed to vary systematically with metallicity, with low-metallicity galaxies exhibiting broader, warmer SEDs and shallower Rayleigh--Jeans slopes, indicative of evolving dust grain properties \citep{remyruyer15,galliano18}, such as dust grain composition and size distribution. At high redshift, such variations remain far less constrained. While there are indications that sub-solar metallicity galaxies at $z \sim 2$ exhibit similarly warm and extended infrared (IR) SEDs \citep{shivaei22a}, a systematic characterization of the SED shape as a function of metallicity, with adequate multi-band coverage and well-constrained Rayleigh--Jeans tails, is still lacking. Constraining these effects is essential for deriving robust dust masses and for accurately quantifying obscured star formation in typical high-redshift galaxies.

The dust-to-gas mass ratio (DtG) of galaxies provides a direct link between metals in the gas phase and those incorporated into dust grains, offering a key diagnostic of dust evolution. In the local Universe, DtG is observed to increase with metallicity, reflecting a transition from stellar dust production to efficient grain growth in the ISM above a critical metallicity $Z_{\rm crit}$ \citep{remyruyer14, devis19, galliano21}. The slope and normalization of the DtG--metallicity relation encode the efficiency of grain growth, the impact of supernova-driven destruction, and the balance between dust production channels \citep[e.g.,][]{popping17}. At $z > 1$, however, constraints on this relation are sparse and largely limited to massive, near-solar metallicity systems or small lensed samples \citep{Saintonge2013,shapley20,Popping2023}. Extending these measurements to sub-solar metallicities at cosmic noon is critical for testing theoretical models, which predict a wide diversity of dust evolution pathways depending on ISM conditions and star formation histories \citep{Popping2022}.

The ALMA Chemical Evolution (ACE) survey is an ALMA Large Program designed to overcome these limitations by providing a consistent and representative dataset of molecular gas, dust, metallicity, and star formation measurements for galaxies at $z = 2.0$--$2.5$. The sample consists of 25 star-forming galaxies drawn from the MOSFIRE Deep Evolution Field (MOSDEF) survey \citep{kriek15}, which provides robust gas-phase metallicities based on multiple rest-frame optical emission lines. ACE combines ALMA CO(3-2) observations probing the molecular gas reservoir with multi-band dust continuum measurements tracing dust emission, complemented by extensive multiwavelength photometry including James Webb Space Telescope (JWST) Near Infrared Camera (NIRCam) and Mid-Infrared Instrument (MIRI) data. By extending to stellar masses and metallicities well below those probed by previous CO surveys, reaching down to $\sim 0.3\,Z_\odot$, ACE explores a previously inaccessible parameter space at cosmic noon, enabling the first systematic study of the interplay between gas, dust, metals, stars, and star formation in typical, low-mass galaxies at this epoch.

In this paper, we introduce the ACE survey and present the sample and its multiwavelength properties. We define the global galaxy parameters used throughout the ACE papers and present the ALMA CO and Band~7 continuum data.
In companion ACE papers, we explore the gas-fundamental metallicity relation \citep{Langan2026}, dust SEDs and temperatures \citep{Popescu2026}, molecular gas scaling relations \citep{Popping2026}, dust mass scaling relations \citep{Solimano2026}, and  dust-to-gas mass ratios \citep{Geesink2026}.
This paper is organized as follows. In Section~\ref{sec:sample}, we present the ACE sample selection, its multi-wavelength coverage, and the available ancillary data. In Section~\ref{sec:alma}, we describe the ALMA observations and data reduction, including the measurements of CO(3-2) line emission and Band~7 dust continuum. In Section~\ref{sec:multiwave}, we detail the derivation of galaxy properties from ancillary data. In Section~\ref{sec:results}, we present the main observational results, including the detection statistics. In Section~\ref{sec:goals}, we summarize the main science goals of the ACE survey and outline the scope of companion papers. Finally, in Section~\ref{sec:summary}, we summarize our main findings. Throughout this work, we assume a \citet{chabrier03} initial mass function (IMF) and adopt a flat $\Lambda$CDM cosmology consistent with \citet{planck2020}, with $H_0 = 67.7\,\mathrm{km\,s^{-1}\,Mpc^{-1}}$, $\Omega_{\mathrm{m}} = 0.31$, and $\Omega_{\Lambda} = 0.69$. We define gas-phase metallicity and the adopted solar reference value in Section~\ref{sec:metallicity}.

\begin{table*}[ht]
\caption{ACE sample and fluxes of CO and continuum observations.}
\centering
\footnotesize
\resizebox{\textwidth}{!}{%
\begin{tabular}{cclcccccc}
\hline\hline
ACE ID & 3DHST-v4 ID & B3 Prog. & $S_{\rm CO(3-2)}$ & $\sigma_{S_{\rm CO(3-2)}}$ & B7 flux & B7 flux error & B6 flux & B6 flux error \\
   &             &          & (mJy\,km\,s$^{-1}$) & (mJy\,km\,s$^{-1}$) & (mJy\,bm$^{-1}$) & (mJy\,bm$^{-1}$) & (mJy\,bm$^{-1}$) & (mJy\,bm$^{-1}$) \\
\hline
$^\dagger$2672 & 2672  & B & 143.5 & 34.3 & \ldots & \ldots & \ldots & \ldots \\
3284  & 3324  & A   & 86.3  & 17.2 & 0.319  & 0.044  & 0.118  & 0.049  \\
3623  & 3626  & A   & 33.2  & 11.1 & 0.019  & 0.036  & 0.019  & 0.012  \\
3692  & 3666  & A+B & 44.4  & 9.9  & 0.259  & 0.071  & 0.040  & 0.030  \\
3773  & 3773  & A   & 37.9  & 12.6 & 0.061  & 0.038  & 0.030  & 0.011  \\
4611  & 4497  & A   & 68.4  & 14.6 & 0.143  & 0.035  & 0.108  & 0.035  \\
5242  & 5094  & B   & 191.2 & 41.3 & 0.326  & 0.049  & 0.074  & 0.047  \\
5814  & 5814  & A   & 119.0 & 19.7 & 0.378  & 0.061  & 0.110  & 0.044  \\
5901  & 5901  & A   & 36.8  & 12.3 & 0.150  & 0.036  & 0.021  & 0.034  \\
6283  & 6283  & A   & 36.9  & 12.3 & 0.030  & 0.035  & 0.007  & 0.011  \\
6750  & 6750  & A   & 34.9  & 11.6 & 0.098  & 0.029  & 0.022  & 0.018  \\
8280  & 8280  & A   & 116.5 & 13.2 & 0.536  & 0.049  & 0.092  & 0.046  \\
8515  & 8515  & A   & 41.1  & 13.7 & 0.010  & 0.033  & 0.046  & 0.012  \\
9393  & 9393  & A   & 38.5  & 9.6  & 0.203  & 0.051  & 0.015  & 0.024  \\
9971  & 9971  & A   & 76.0  & 19.2 & 0.193  & 0.053  & 0.098  & 0.027  \\
13895 & 13701 & B   & 116.1 & 22.4 & 0.489  & 0.057  & 0.047  & 0.053  \\
13899 & 13296 & B   & 88.0  & 18.1 & 0.227  & 0.051  & 0.152  & 0.052  \\
17482 & 16594 & A   & 62.9  & 13.1 & 0.137  & 0.035  & 0.032  & 0.032  \\
19876 & 19013 & A   & 58.5  & 14.4 & 0.265  & 0.043  & 0.020  & 0.036  \\
20331 & 19439 & A   & 31.7  & 10.6 & 0.036  & 0.039  & 0.045  & 0.020  \\
20891 & 19985 & A   & 59.9  & 12.9 & 0.272  & 0.035  & 0.104  & 0.019  \\
21955 & 21955 & A   & 67.2  & 16.1 & 0.165  & 0.039  & 0.064  & 0.033  \\
22193 & 22193 & A   & 33.8  & 11.3 & 0.051  & 0.035  & 0.023  & 0.036  \\
24020 & 24020 & A   & 32.9  & 11.0 & 0.008  & 0.031  & 0.017  & 0.011  \\
24763 & 24763 & A+B & 66.7  & 13.0 & 0.354  & 0.050  & $-$0.041 & 0.048 \\
25229 & 25229 & A   & 40.9  & 12.3 & 0.037  & 0.034  & 0.027  & 0.014  \\
\hline
\end{tabular}%
}
\tablefoot{Columns are: 1) ID used in ALMA observations, 2) object's ID in the 3D-HST v4.0 catalog \citep{skelton14}, 3) program ID from which the CO data is taken: A: 2024.1.00534.L (ACE); B: 2018.1.01128.S \citep{Sanders2023}. In some cases, data from both programs are combined. 4--5) CO(3-2) line flux and uncertainty. For analysis details, see \citet{Langan2026}. 
The line flux is derived from the Gaussian fit to the CO line in the 1D spectra after random resampling of the observed spectrum, extracted at the peak pixel of the moment 0 map. The error is defined as the larger of the standard deviation from the resampling or the RMS within the Gaussian FWHM. 
6--9) Continuum flux in Band 7 (from ACE) and Band 6 (from \cite{shivaei22a}, program 2019.1.01142.S). For analysis details, see \citet{Popescu2026}.
The continuum flux values are derived from the continuum peak flux density of the tapered image and its rms. If the source is not detected with greater than $3\sigma$, this is the value of the flux at the source coordinates. 
Those with S/N$<2$ in continuum and S/N$<3$ in CO are set to $-99$. \\
$^{\dagger}$ This source was not observed in Band 6 or 7 as part of ACE, but since it has CO observations from program 2018.1.01128.S and metallicity from MOSDEF, it is included in the analyses.}
\label{tab:ace_sample}
\end{table*}

\section{Sample} \label{sec:sample}
The ACE sample is based on a Cycle-9 ALMA Band~6 program \citep[\#2019.1.01142.S;][]{shivaei22a} sample selected from the near-IR spectroscopic survey of MOSDEF \citep{kriek15}. MOSDEF is a Keck/MOSFIRE survey of a H (F160W)-band selected sample of $\sim1500$ galaxies at $z = 1.4-3.8$. MOSFIRE is a near-IR multi-object spectrometer with a resolving power of $R\sim3300$ for a 0.7$''$ slit width \citep{mclean10,mclean12}.
The MOSDEF observations were taken in Y, J, H, K bands covering all of the strongest emission features between rest-frame 3700 and 6800\,$\AA$ including H$\alpha$, H$\beta$, [O{\sc ii}], [O{\sc iii}], [N{\sc ii}], and [S{\sc ii}]. This dataset provides spectroscopic redshifts, gas-phase metallicities, and dust-corrected SFRs. Additionally, the uniqueness of this dataset in the era of JWST/NIRSpec is the small slit-loss corrections given that the sizes of galaxies are significantly large compared to the NIRSpec MSA microshutters sizes ($0.2''$ slit width, \citealt{Jakobsen2022,Ferruit2022}), but are more comparable to the MOSFIRE slit widths of $0.7''$ (even considering the typical seeing of $0.5-1.0''$ during MOSDEF observations \citealt{kriek15}).

The original Band~6 sample was selected in the COSMOS field to satisfy these criteria: (1) redshift of $2.0 < z < 2.5$, where ALMA Bands~3 and~7 capture the CO(3-2) and dust Rayleigh--Jeans (RJ) emission, respectively, and JWST/MIRI traces the 7.7\,$\mu$m polycyclic aromatic hydrocarbon (PAH) emission; (2) $> 3\sigma$ detections in the H$\alpha$, H$\beta$, [O{\sc iii}], and [N{\sc ii}] lines, to ensure robust estimates of SFR and metallicity; (3) no evidence of AGN activity, based on X-ray emission, mid-IR (IRAC) colors, and optical line ratios \citep{coil15}. The sample is further refined for ACE by excluding two of the original Band~6 targets due to either disjoint dust/optical morphologies or an uncertain metallicity estimate (0.5\,dex uncertainty), leaving a total of 25 targets \citep{shivaei22a}. To enhance the science return and efficiency of ACE, we combined our sample with robust CO detections from a Cycle-8 ALMA program (\citealt{Sanders2023}, \#2018.1.01128.S), which was also based on the MOSDEF parent sample. In that program, 10 sources were observed in Band~3 for CO(3-2) emission, and 3 of them were robustly detected (signal-to-noise ratio of $S/N > 4$). These three were only observed in Band~7 as part of the large program. Additionally, we added another target (ID 2672) from the \cite{Sanders2023} program with significant CO detection to our CO sample (this object was not in the original ACE sample due to the lack of {\Hb} line detection). Consequently, this object does not have Bands 6 or 7 continuum observations and is only used in the CO analyses (shown by a $\dagger$ sign in Table~\ref{tab:ace_sample}).

\begin{figure*}[ht]
        \centering
        \includegraphics[width=.9\textwidth]{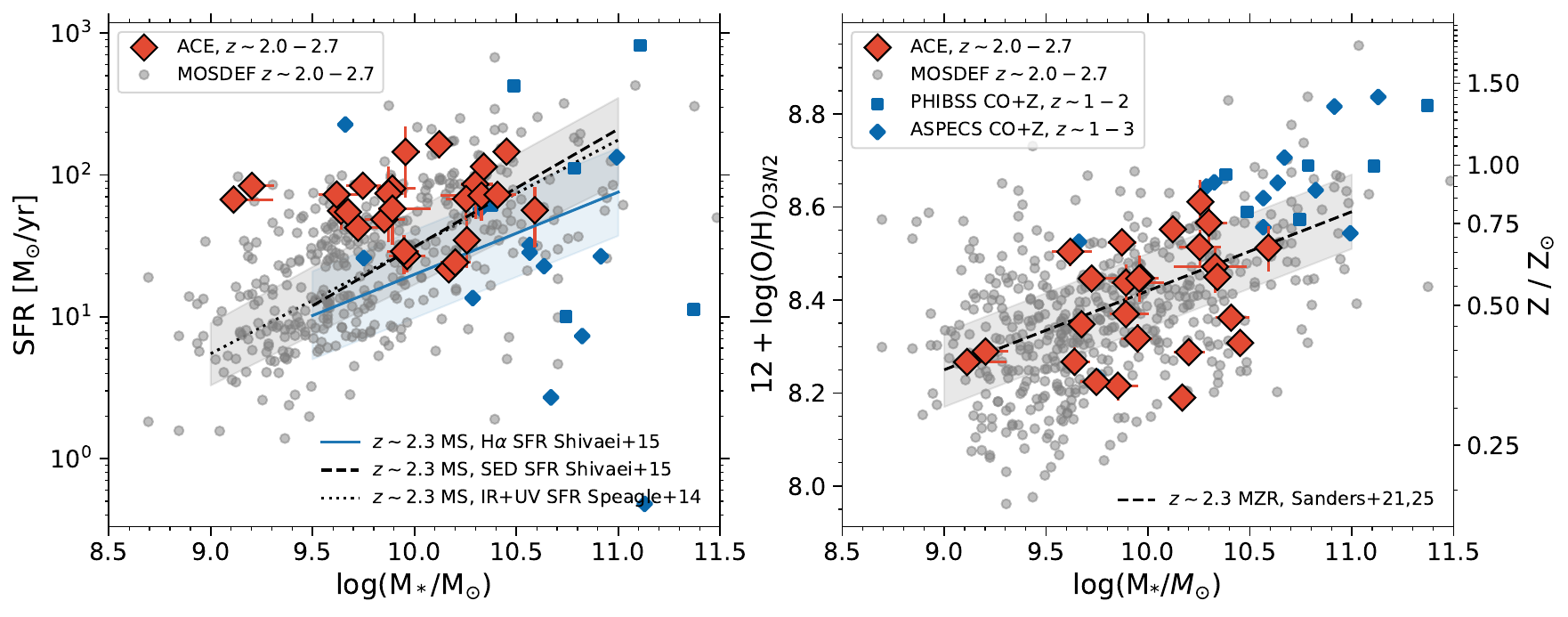} 
    \caption{\label{fig:MS_MZR}
    Star forming main sequence (left) and mass-metallicity relation (right) for the ACE sample (red diamonds), and its parent sample of the MOSDEF galaxies at $z=2.0-2.7$ (grey circles). SFRs are derived from the attenuation-corrected H$\alpha$ luminosities using Balmer decrement (Section~\ref{sec:sfr}; a version of this plot with SED-inferred SFRs is shown in Appendix~\ref{sec:otherMS}), stellar masses are derived from SED fitting (Section~\ref{sec:sed}), and metallicities are from multiple optical emission lines (Section~\ref{sec:metallicity}). The selection criterion of detection in multiple optical emission lines has biased the ACE sample to higher SFRs at a given stellar mass compared to the main-sequence relation.    
    For comparison, we show data from two large CO surveys at redshifts similar to those of the ACE sample, PHIBSS \citep{tacconi13} and ASPECS \citep{walter16}, for which metallicity measurements and JWST photometric data are available. The stellar masses and metallicities have been recalculated using the same methodology as for the ACE sample. However, unlike for ACE galaxies, the SFRs are SED-inferred due to lack of {\Ha} and {\Hb} measurements (Figure~\ref{fig:other-ms} in Appendix~\ref{sec:otherMS} shows the relation using SED-inferred SFRs for all samples).
    It is clear that ACE galaxies extend the existing samples of normal (i.e., not SMG, ULIRG, etc) unlensed galaxies to much lower mass and metallicity.
    For reference, various main-sequence (Section~\ref{sec:MS}) and mass-metallicity relations are shown with lines and shaded regions indicating their scatter. 
    }
\end{figure*}

The final ACE sample consists of 26 galaxies (25 with continuum observations), as reported in Tables~\ref{tab:ace_sample} and \ref{tab:ace_sample_props} and shown in Figure~\ref{fig:MS_MZR}. The ACE galaxies all have stellar masses of $M_{\star} \leq 10^{10.5}\,\rm{M}_\odot$, with SFRs ranging from $\sim 20$ to $\sim 200\,\rm{M}_\odot\,\rm{yr}^{-1}$ and metallicities in the range $8.2 < 12 + \log{\mathrm{(O/H)}} < 8.6$. The galaxies are located on or above the main-sequence and nicely sample the mass-metallicity relation from the MOSDEF parent sample.  In Figure~\ref{fig:MS_MZR}, we show the samples from the two main existing CO surveys of normal main-sequence galaxies, PHIBSS and ASPECS, at similar redshifts with available metallicity measurements, for comparison with the ACE sample in the MZR plane. It becomes immediately apparent from Figure~\ref{fig:MS_MZR} that the ACE galaxies probe a region of stellar mass and metallicity space that has not previously been covered by (unlensed)\footnote{We did not include lensed galaxies from the literature due to uncertainties in lensing models and different selection functions.} galaxies.

For every ACE target there exists broad multiwavelength coverage, including HST, JWST NIRCam and MIRI 7.7$\mu$m coverage from COSMOS-Web \citep[JWST 1727,][]{casey23,Shuntov2025}, NIRCam and partial longer-wavelength MIRI coverage from PRIMER (JWST 1837) and COSMOS-3D (JWST 5893), Spitzer/MIPS 24\,$\mu$m \citep{sanders07}, Herschel/PACS\footnote{We did not consider Herschel/SPIRE data due to significant confusion for these relatively low mass and less IR luminous galaxies.} \citep{lutz11}, and ALMA/band-6 data, which together characterize the behavior of UV-optical stellar and nebular emission and the mid-to-far-IR dust emission.

\begin{figure*}[ht]
        \centering
        \includegraphics[width=.9\textwidth]{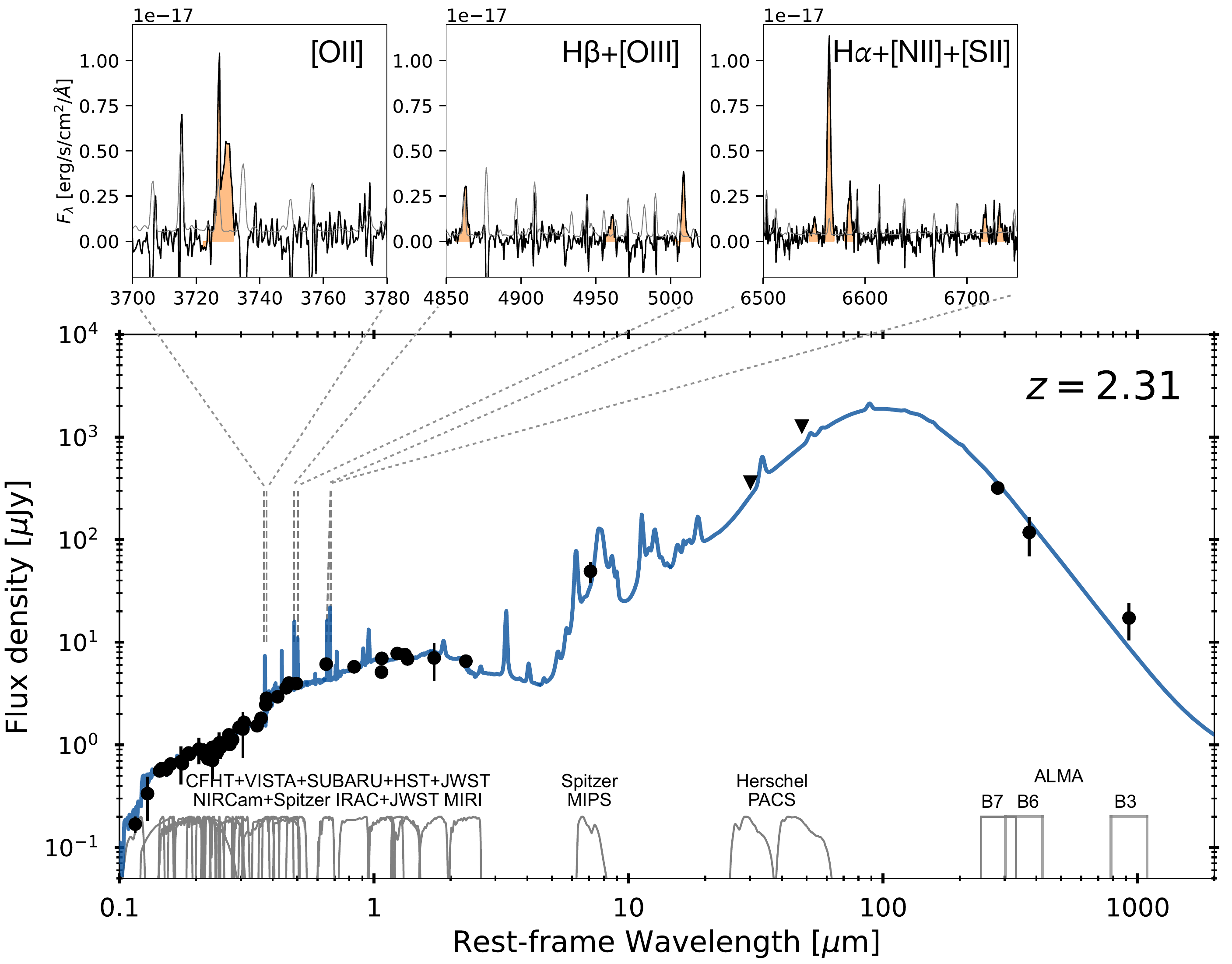} 
    \caption{
    Best-fit SED of one of ACE sources (ID 3324, bottom) and its Keck/MOSFIRE 1D spectrum in J, H, and K bands (top). The filter emission curves used in the SED fitting are shown in the bottom, along with the observed photometry. The Herschel PACS data are shown as 1$\sigma$ upper limits (triangles) due to large errors. Also, IRAC Ch4 photometry is not shown as it is superseded by the higher S/N JWST MIRI F770W photometry. 
    In the spectra on top, the main optical emission lines are shown in orange: [O{\sc ii}]$\lambda\lambda3727,3730$, {\Hb}, [O{\sc iii}]$\lambda\lambda4960,5008$, {\Ha}, [N{\sc ii}]$\lambda\lambda6550,6585$, and [S{\sc ii}]$\lambda\lambda6718, 6733$.
    }
    \label{fig:sed}
\end{figure*}

\section{ALMA data and measurements} \label{sec:alma}

The ACE observations were designed to detect $^{12}$CO\ $J_{\rm up}$=3$\rightarrow$2 transition at $\nu_{\rm rest} = 345.8$\,GHz, in ALMA Band 3 at the redshift of the targets, and the observed-frame $\sim 870\,\mu$m dust continuum in ALMA Band 7. CO(3-2) is the optimal molecular gas mass diagnostic line that is observable at $2.0 < z < 2.5$ with current facilities. Higher transitions of CO are susceptible to larger excitation correction uncertainties, while CO(2–1) was not observable with ALMA at these redshifts at the time of requesting the data\footnote{As of Cycle 13, it is possible with ALMA Band 2 to observe CO(2-1) at these redshifts, however about $4\times$ more expensive than CO(3-2).}, and detecting CO(1–0) would require over a day of on-source time per target.

The planned observations probed the CO(3-2) emission of $z\sim2$ galaxies in an unexplored territory, as the validity of the existing CO luminosity to molecular gas mass scaling relations \citep[e.g.,][]{tacconi18} had not been tested at low metallicities and stellar masses.
We adopted multiple approaches to predict the molecular ($H_2$) masses, and hence CO luminosity for our sample. However, the discrepancies based on different scaling relations in the literature are substantial. In fact, the estimated $H_2$ masses for our targets differ by more than a factor of 3 when adopting the scaling relations of \citet{Sanders2023}, \citet{liu19}, and \citet{tacconi18}. Therefore, instead of adopting uncertain relations in this regime, our sensitivity calculations were guided by the detections of CO(3-2) emission in a stack of six individually CO-undetected galaxies from \citet{Sanders2023}. These 6 individual objects are part of our sample, with an average metallicity of {\OH} = 8.25. The cycle-6 (pre-ACE) observations had $\sim 3-6$ times lower sensitivity than requested in ACE. From their stack (S/N$=$4.3) we expected an average peak flux density of 0.79\,mJy per beam for our targets. Using those observations as a guide, we requested a sensitivity of 0.13\,mJy per beam per channel of 50\,km/s in Band 3 to securely detect our targets with a S/N of at least 6 individually.

The details of the data reduction and flux measurements for both the CO(3-2) line and the Band 7 continuum are reported in \citet{Langan2026} and \citet{Popescu2026}, respectively. Briefly, we utilize the calibrated measurement sets provided by the European ALMA Regional Centre via the calMS service \citep{Petry2020}. Imaging is performed using the \textsc{Common Astronomy Software Applications} (\textsc{CASA}, v6.5.21; \citealt{CASA}) software package. For both the Band 7 continuum and the CO(3-2) line emission, we construct images using the \texttt{tclean} task. We generate initial images using natural weighting, alongside additional versions applying 1 and $2\,\mathrm{arcsec}$ $uv$-tapers. Further details regarding the resulting beam sizes and sensitivities are provided in \citet{Langan2026} and \citet{Popescu2026}.

Because our primary objective is to capture the total integrated emission of the targets, we measure the Band~7 continuum flux from the highest-resolution image (selected from the naturally weighted, $1\,\mathrm{arcsec}$, or $2\,\mathrm{arcsec}$ tapered maps) in which the source remains spatially unresolved (see \citealt{Popescu2026} for more extensive tests justifying this approach). The CO(3-2) line emission is measured via an iterative procedure. We first collapse the data cube over a velocity range corresponding to $\pm\mathrm{FWHM}_{\mathrm{H}\alpha}$ (full width at half maximum of {\Ha}; \citealt{kriek15,reddy15}) centered on the systemic redshift to create an intensity map. From this preliminary map, we locate the emission peak and extract a 1D spectrum, which is then fitted to determine the $\mathrm{CO}$ line width. This line width is subsequently used to define the optimized velocity interval ($\pm\mathrm{FWHM}_{\mathrm{CO}}$) for the final moment-0 map. As a final step, the definitive CO(3-2) spectrum is extracted from the cube at the peak pixel location of the moment-0 map. Similar to our continuum approach, the final $\mathrm{CO}$ spectra are measured using the highest-resolution image configuration where the moment-0 emission remains unresolved.

Out of the 23 CO observations in ACE, 14 were detected in CO(3-2) at S/N>3 (61\% success rate); incorporating the CO observations from previous cycles \citep{Sanders2023}, that adds to total of 17 CO(3-2) detections. Of the 25 Band 7 continuum observations, 17 are detected at S/N$>3$ (68\% success rate). Four of the ACE sources were previously detected in Band 6 as well. Three sources are marginally detected in Band 3 continuum with S/N >2.

\section{Multi-wavelength ancillary data and Measurements} \label{sec:multiwave}

One of the unique features of ACE is the wealth of ancillary data available for the sample. ACE sample was selected based on the MOSDEF survey to have robust detections in primary rest-frame optical emission lines from the near-IR MOSFIRE spectrograph (Section~\ref{sec:sample}). The optical lines are used to derive redshifts, ionized gas properties, such as metallicity, and instantaneous SFR (Sections~\ref{sec:metallicity} and \ref{sec:sfr}).
As the sources are located in the COSMOS field, they have various photometric data from ground-based telescopes Subaru, VISTA, and CFHT, as well as HST, Spitzer, Herschel, and JWST. The optical to near-IR data used for ACE are from the compilation of COSMOS-Web survey \citep[see also \citealt{Weaver2022}]{casey23, Shuntov2025}. The mid- and far-IR data of Spitzer and Herschel are from the compilation of \cite{shivaei17,shivaei22a}. ALMA continuum data of Band~7, 6, and 3 are added following the new reduction and photometry of \citet{Popescu2026}.
These photometric data, combined with ALMA band 7, 6, and 3, are used to derive stellar population properties of ACE galaxies (Section~\ref{sec:sed}).

We also explored the 21\,$\mu$m MIRI coverage of the ACE sample from various public programs in the COSMOS field listed in \citet[][JWST 2417 and 5893]{perez-gonzalez26}. The data is taken from the JWST Rainbow repository and is reduced coherently as explained in \cite{perez-gonzalez24,perez-gonzalez26}. The JWST Rainbow Pipeline adopts the official JWST pipeline with added steps for background homogenization of the MIRI images. We only explored MIRI F2100W band as it traces the hot dust and PAH 6-7\,$\mu$m features at the redshift of ACE sample, providing interesting additional information on dust species in these galaxies (Section~\ref{sec:morphology}.

\subsection{SED fitting} \label{sec:sed}
We use the \texttt{Prospector} SED fitting code \citep{johnson21}, a Bayesian forward modeling code that adopts Monte Carlo sampling of the parameter space. We fit the models to rest-frame UV to far-IR data, using Flexible Stellar Population Synthesis (FSPS) library \citep{conroy13} and dust templates of \cite{draineli07}. The redshift of each target is known from the spectroscopic data. We follow the code assumptions and prior setup of \cite{shivaei24a}, which is briefly discussed here. An example of the best-fit SED and available photometry for one of the ACE galaxies (ID V4 3324) is shown in Figure~\ref{fig:sed}.

\paragraph{IR SED} The \cite{draineli07} model assumes a mixture of amorphous silicate and carbonaceous grains, with a grain-size distribution selected to reproduce the wavelength dependence of the Milky Way (MW) extinction curve. The relative silicate and carbonaceous abundances are constrained by observations of gas-phase depletions in the ISM. The dust-heating radiation field is taken to have the fixed spectral shape of the local interstellar radiation field \citep{mathis83}, scaled by a dimensionless factor $U$. The free parameters in our fits are: $q_{\rm PAH}$, the mass fraction of grains in PAHs containing fewer than $10^3$ carbon atoms; $U_{\mathrm{min}}$, the intensity of the diffuse ISM radiation field heating the dust; and $\gamma$, the fraction of dust mass exposed to a power-law distribution of starlight intensities between $U_{\rm min}$ and $U_{\rm max}$ (with $1-\gamma$ representing the fraction exposed to $U_{\rm min}$). We assume flat priors of $U_{\rm min}=0.1-15$, $\gamma=0.001 - 0.15$, and $q_{\rm PAH}=0.4-4.6$. The latter is the range of $q_{\rm PAH}$ that the \cite{draineli07} models are specifically constrained for. 

\paragraph{UV-optical SED}
We assume a delayed-$\tau$ star formation history with a flat age prior between 1\,Myr to the age of the universe at the redshift of the galaxy. The stellar metallicity is left free with a flat prior between $\log(Z/Z_{\odot}) = -2.5$ and 0.19. We assume a nebular emission model with flat priors in logarithmic space of gas-phase metallicity at $-2.0$ to 0.5 (relative to solar metallicity) and ionization parameter at $-4$ to $-1$. We include IGM absorption \citep{madau95} with the optical depth scaling factor as a free parameter with a Gaussian prior centered at 1.0 and $\sigma=0.3$. Following \cite{shivaei26}, we adopt a modified attenuation curve with a UV-optical slope as a free parameter and independent from the UV bump amplitude, as parameterized in \cite{salim18,noll07}. The curve has a flat prior of $-0.6$ to 0.3 for the multiplicative coefficient of the slope of the \cite{calzetti00} curve. Throughout the paper, we adopt the ``surviving'' stellar mass (not ``formed'' mass), which takes into account mass losses throughout the evolution of stars in the galaxy via AGB winds, supernovae, etc. 

We do not include AGN emission, as ACE targets are specifically selected to be star-forming, excluding AGN based on multiple diagnostics, including mid-IR (IRAC) colors, X-ray emission, and BPT classifications. The only galaxy that may have a buried obscured AGN is ID 8280, however, with only MIPS 24\,$\mu$m data the origin of its obscured emission (AGN or obscured SF) is unclear \citep[see][]{shivaei22a}, and requires more investigation with MIRI multi-band mid-IR coverage. Given the low metallicity of most of the sample, including AGN models would lead the fitting code to artificially increase the mid-IR AGN luminosity fraction to reproduce the warm dust emission. This bias is particularly pronounced in the absence of full MIRI coverage, resulting in incorrect stellar mass estimates for the sample. We will compare the SED-inferred SFRs with and without AGN models in Section~\ref{sec:sfr}, showing the ones with AGN models included are in less agreement with the independently derived {\Ha} SFRs.

\begin{figure}[ht]
        \centering
        \includegraphics[width=.9\columnwidth]{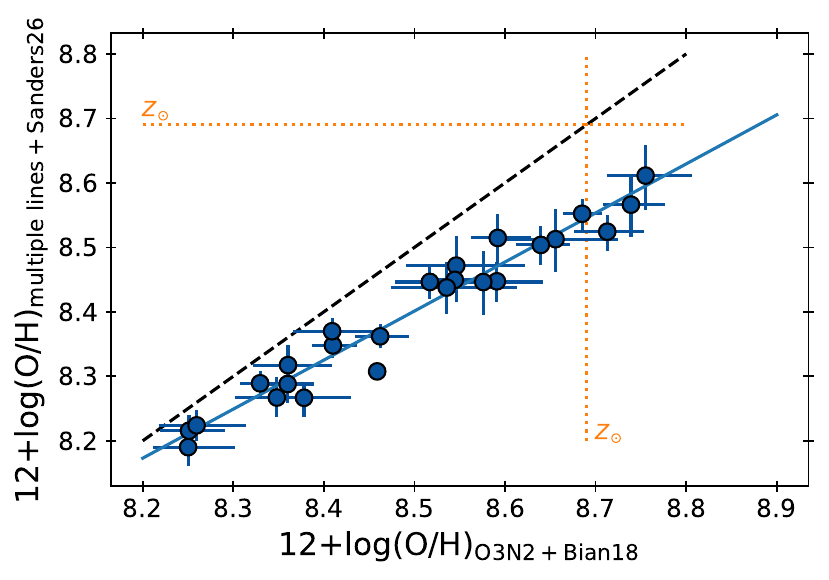} 
    \caption{Comparison of the two metallicities based on the pre- and post-JWST calibrations for the 25 ACE galaxies from \cite{shivaei22a}. The horizontal axis shows the pre-JWST metallicities from \citealt{shivaei22a}, based on the O3N2$=$([O {\sc iii}]$\lambda5007$/H$\beta$)/([N {\sc ii}]$\lambda 6584$/H$\alpha$) line ratio and the \cite{bian18} calibration. The vertical axis shows the currently adopted metallicities of ACE from different line ratios (depending on availability) and calibrations of \cite{sanders26} using a minimization method (see text). The best-fit line is shown in blue, which is $y=0.76\times x + 1.94$.
    The main difference occurs at the high-metallicity end, where the full ACE sample is sub-solar with the new calibrations (highest metallicity is 8.64). Throughout the ACE papers we recalculate all the cosmic noon literature comparison metallicities using the same method and calibration as applied to the ACE galaxies.}
    \label{fig:Z_comp}
\end{figure}

\subsection{Metallicities} \label{sec:metallicity}
Metallicity is defined as gas oxygen abundance, $12+\log({\rm O/H})$. Throughout the paper, solar metallicity is adopted to be 8.69 \citep{asplund09}.
All of the ACE sample have the main strong optical lines, H$\alpha$, H$\beta$, [O {\sc iii}]$\lambda5008$, [N {\sc ii}]$\lambda6585$, to derive metallicites, except for one target added from previous programs (ID 2672) that is undetected in H$\beta$. In addition to these lines, several targets also have other lines detected, such as [O {\sc ii}]$\lambda\lambda3727,3730$, [S {\sc ii}]$\lambda\lambda6718,6733$, [Ar {\sc iii}]$\lambda7137$, [Ne {\sc iii}]$\lambda3870$, H$\gamma$. Line fluxes and uncertainties are adopted from the MOSDEF survey. We refer to \cite{reddy15} and \cite{kriek15} for details on data reduction and line measurements.

To take advantage of the full range of available lines for each source, and to avoid biases related to certain abundances such as N/O, we use all available line ratios and adopt the calibrations of \cite{sanders26}. Final metallicities are calculated via a $\chi^2$ minimization over multiple line ratios simultaneously \citep[e.g., see][]{sanders21}. We include both the measurements uncertainty from line flux measurements, and the calibration model scatter ($\sigma_{R,fit}$ in \citealt{sanders26}). The new metallicities are reported in Table~\ref{tab:ace_sample_props}. We note that the new strong-line calibrations of \cite{sanders26}, based on the $T_e$ method, are only extrapolated beyond metallicity of $12+\log({\mathrm O/H})=8.6$ and yield systematically lower metallicities at high metallicities compared to the commonly adopted pre-JWST calibrations \citep[e.g.,][see Figure~\ref{fig:Z_comp}]{bian18}. This results in all of ACE sample falling below solar metallicity. A comparison of the previously measured metallicities \citep[from][]{shivaei22a} using the \cite{bian18} calibrations based on a local sample of high-redshift analogs with the new metallicities is shown in Figure~\ref{fig:Z_comp}. The shift in the calibration is characterized by a linear function:
\begin{equation}
    Z_{\rm new} = 0.76\times Z_{\rm Bian18} + 1.94,
\end{equation}
in which $Z$ is $12+\log{\rm O/H}$, and $Z_{\rm new}$ corresponds to the calibrations of \citet{sanders26} using multiple emission lines. Given this significant shift in high-redshift calibrations, throughout the ACE papers we recalculate the literature metallicities using the same method and calibration applied to the ACE galaxies.

\begin{figure}[ht]
        \centering
        \includegraphics[width=.9\columnwidth]{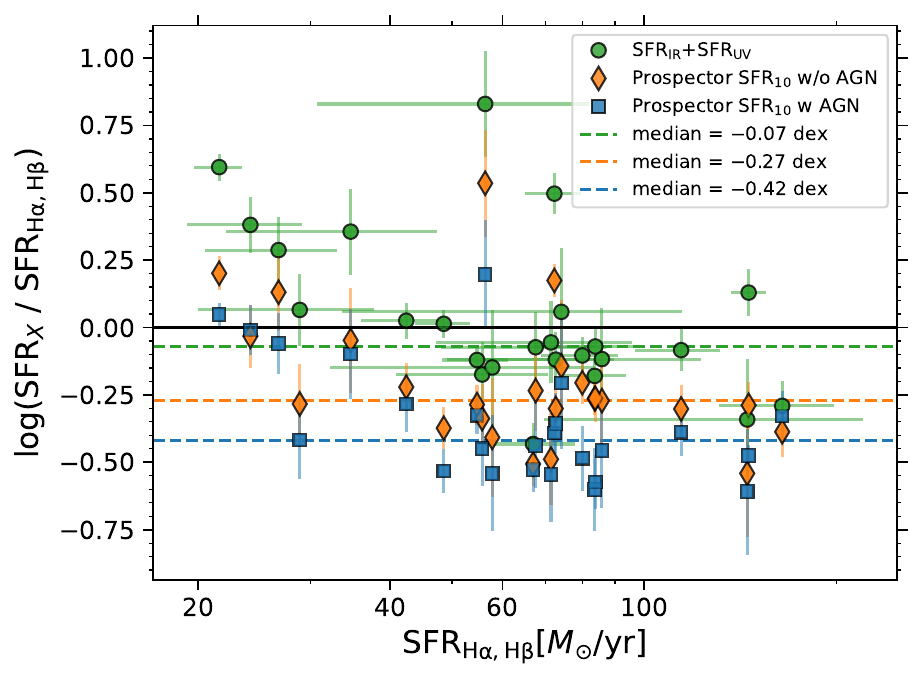} 
    \caption{Comparison of different SFR estimates. Horizonal axis shows SFR$_{{\rm H\alpha,H\beta}}$, dust-corrected H$\alpha$ SFR (assuming two different metallicity-dependent conversions, see text). Vertical axis shows the ratio of other SFR estimates to SFR$_{{\rm H\alpha,H\beta}}$: SFR$_{\rm UV}$ and SFR$_{\rm IR}$ are derived from the best-fit SEDs, and SFR$_{10}$ is the 10\,Myr averaged SED-inferred SFR with and without AGN models. Dashed lines show the median values of the ratios corresponding to the same colors.
    SFR$_{{\rm H\alpha,H\beta}}$ and SFR$_{\rm UV}$+SFR$_{\rm IR}$ are in good agreement, with the exception of one galaxy, which likely has very warm dust due to recent starburst activity or buried AGN (see text).}
    \label{fig:SFR_comp}
\end{figure}

\subsection{SFR diagnostics} \label{sec:sfr}
ACE galaxies have multiple tracers of SFR, including Balmer decrement dust-corrected H$\alpha$ SFRs, IR and UV SFRs, and SFRs from full (energy-balanced) SED fitting. In this section, we compare the various SFR indicators and discuss their differences.

The optical nebular Hydrogen lines, such as H$\alpha$, are considered the gold standard for instantaneous SFR measurement because they 1) trace the star formation activity over very short time scales of primarily $<10-20$\,Myr (as they trace the ionizing UV emission from the most recently born stars; \citealt{kennicutt98}), and 2) are less affected by dust attenuation compared to direct UV diagnostics. Particularly, with availability of another Balmer line (such as H$\beta$), H$\alpha$ can be robustly corrected for dust attenuation \citep{reddy15, shivaei15a, shivaei15b,shivaei16}\footnote{Although recent studies of higher redshift galaxies indicate that Balmer decrement might not fully capture the total dust attenuation in the nebular regions \citep{reddy26}.}.
We calculated H$\alpha$, H$\beta$ SFRs following the procedure of \citet[][Section 2.4.1]{shivaei22a}, as follows.
We assume the \citet{cardelli89} Milky Way extinction curve to derive dust attenuation of H$\alpha$ line using the Balmer decrement \citep{reddy20,rezaee21}.
To take into account the metallicity dependence of the H$\alpha$ luminosity to SFR conversion \citep{theios19}, we adopt two different conversion factors for galaxies above and below the gas metallicity of $12+\log(\rm O/H) \sim 8.6$ (assuming the stellar and ISM gas metallicity are linearly correlated; \citealt{gallazzi05}). For the high-metallicity galaxies, we adopt the conversion factor of \citet{hao11} converted to a \citet{chabrier03} IMF, $\log({\mathrm{SFR})=\log{\rm(L(H\alpha)}})-41.26$, where SFR and L(H$\alpha$) are in $M_{\odot}$/yr and erg/s units, respectively. For the lower-metallicity galaxies, we adopt the conversions derived from the \citet{bc03} stellar population models with constant star formation over 100\,Myr and $Z = 0.004$ from \citet{reddy18b} and \citet{theios19}, adjusted to the \citet{chabrier03} IMF: $\log({\mathrm{SFR})=\log{\rm(L(H\alpha)}})-41.41$. For reference, the \citet{kennicutt98} and \citet{kennicutt12} conversion constants for the same IMF assumed here, are $-41.30$ and $-41.26$, respectively.

Figure~\ref{fig:SFR_comp} shows the comparison of the Balmer decrement dust-corrected SFR$_{{\rm H\alpha,H\beta}}$ with IR$+$UV SFR and SED-inferred SFRs. SFR$_{\rm UV}$ and SFR$_{\rm IR}$ are derived from the luminosity at 1600\,$\AA$ and $8-1000\,\mu$m of the best-fit \texttt{Prospector} SED fits (Section~\ref{sec:sed}), using SFR-luminosity calibrations of \citet{kennicutt12}. SFR$_{10}$ is the 10\,Myr averaged SFR from the \texttt{Prospector} SED fits. We show these SED SFRs for both the fits with and without AGN. The fits with AGN (which are not considered in the ACE analyses, see Section~\ref{sec:sed}) are in least agreement with the SFR$_{{\rm H\alpha,H\beta}}$ -- another reason for not considering the AGN models in the fits for these galaxies. While the fits without AGN are in better agreement, still there is a large descripancy (average of 0.3\,dex). This discrepancy may be partly attributable to the manner in which the SFR is inferred by {\texttt{Prospector}}. Although the fits assume UV--IR energy balance, a substantial fraction of the IR emission may be attributed to dust heated by old stellar populations, which do not contribute to the recent SFR \citep{leja22}. Such a contribution is likely to be less significant in the actively star-forming galaxies in the ACE sample, and the SFRs inferred by \texttt{Prospector} may therefore be systematically underestimated. SFR$_{\rm UV}+$SFR$_{\rm IR}$ and SFR$_{{\rm H\alpha,H\beta}}$ are in good agreement with each other within uncertainties\footnote{The IR and UV SFR uncertainties are likely underestimated, as they depend on the initial priors and the assumed IR models.} (median offset of 0.07 dex). The good agreement between dust-corrected H$\alpha$ and IR$+$UV SFRs at cosmic noon has been shown previously as well \citep{shivaei16}. 

The object with the most discrepant SFR$_{{\rm H\alpha,H\beta}}$ and SFR$_{\rm UV}$+SFR$_{\rm IR}$ (0.83\,dex; ID 8280) is a very IR-bright galaxy with detections in PACS 100, 160, and ALMA band 7 and 6. The galaxy has high metallicity (12+$\log$(O/H)$=$8.55), but a young age of 100\,Myr. Furthermore, it shows an elevated 24-160$\,\mu$m emission compared to its 250-1200\,$\mu$m, indicating the presence of warm dust, which can be either due to a recent starburst given its young age, a very dusty star-forming galaxy where the Balmer lines are primarily sensitive to the optically thin OB associations \citep[e.g.,][]{reddy26}, or an obscured AGN, where the infrared emission is dominated by the AGN emission but the optical lines are obscured (which can be tested with more mid-IR coverage with MIRI). Regardless of the origin, this galaxy clearly has a high obscured fraction and its optical nebular emission lines may be partially optically-thick, and hence, the observed fluxes are underestimating the total emission even after dust correction.

For the remainder of this paper, as well as in most other ACE papers, we adopt SFR$_{{\rm H\alpha,H\beta}}$ as our fiducial SFR indicator. However, when comparing with datasets that lack {\Ha} and {\Hb} measurements, we adopt SFR$_{\rm UV}+$SFR$_{\rm IR}$ for consistency.

\subsection{Main-sequence of star-forming galaxies}\label{sec:MS}
The star-forming main sequence is one of the most fundamental scaling relations in galaxy evolution, linking stellar mass and star formation rate, and providing insight into the long-term growth of galaxies. In addition to its physical significance, the main sequence serves as a practical reference throughout the ACE analyses and is shown in Figure~\ref{fig:MS_MZR}. In particular, several companion papers use a galaxy's offset from the main sequence ($\Delta$MS) to quantify its level of star-forming activity relative to the typical galaxy at the same stellar mass, providing a common framework for interpreting molecular gas and dust properties of the ACE (and literature) galaxies. 

For the MOSDEF survey, \citet{shivaei15b} derived a main-sequence relation for star-forming galaxies at $2.0 < z < 2.7$ based on dust-corrected H$\alpha$ star formation rates. Since that work, the ACE galaxies have been reanalyzed using updated \texttt{Prospector} SED modeling that incorporates the extensive JWST photometric data now available for the sample, as well as the IR constraints including the new ALMA data. This leads to systematic downward shift in the inferred stellar masses relative to those adopted by \citet{shivaei15b} of on average a factor of 1.8.
This shift can be due to better photometric coverage in the new fits (particularly the IR data points) as well as different SED fitting codes with different underlying assumptions. We note that here we adopted the surviving stellar mass, which accounts for stellar mass loss over a galaxy's lifetime, instead of the total mass of stars formed (integrated over the star formation history).
To maintain consistency with the H$\alpha$-based star formation rates used throughout the ACE survey, we adjust the normalization of the \citet{shivaei15b} main sequence to account for the typical stellar-mass offset between the original \citet{shivaei15b} analysis and the updated ACE measurements. The resulting relation, applicable for galaxies at $z\sim 2.3$,  adopted throughout most of the ACE papers is 
\begin{equation} 
\log \left( \frac{\mathrm{SFR}_{\mathrm{H}\alpha}}{M_\odot\,\mathrm{yr}^{-1}} \right) = 0.58 \, \log \left( \frac{M_\star}{M_\odot} \right) - 4.50.
\end{equation}

This relation and its intrinsic scatter of 0.31\,dex are shown in Figure~\ref{fig:MS_MZR}. The majority of our galaxies (76\%, 19 galaxies) lie above the scatter of this main-sequence relation. However, this definition changes by assuming different SFR indicators and main-sequence definisions.
Assuming SFR(IR)+SFR(UV) and the \citet{speagle14} respective (based on SFR(IR)+SFR(UV)) main-sequence relation and scatter (shown in Figure~\ref{fig:MS_MZR}), 64\% (16 galaxies) are above the main-sequence. However, assuming the SED-inferred SFRs and the SED-based main-sequence of \citep{shivaei15} (and scatter), only 5 (20\%) are above the main sequence. The latter is shown in Figure~\ref{fig:other-ms}.

\subsection{Dust and molecular gas masses}
The derivation of dust and molecular gas masses is presented in detail in the companion papers of \citet{Solimano2026} and \citet{Langan2026}, respectively. Dust masses are estimated from the ALMA continuum measurements assuming optically thin emission and modeling the far-infrared spectral energy distribution with a modified blackbody. The adopted dust temperatures and emissivity indices are informed by the combined ALMA Band~3, 6, and 7 continuum data together with available Herschel measurements, as described by \citet{Popescu2026}. The resulting dust masses provide one of the first systematic measurements of the dust content of sub-solar metallicity galaxies at cosmic noon \citep{Solimano2026}.

Molecular gas masses are derived from the observed CO(3-2) luminosities. These are converted to CO($1$--$0$) luminosities using an excitation correction appropriate for main-sequence galaxies at $z\sim2$ \citep[][$r_{31} = L'_{\mathrm{CO}(3-2)}/L'_{\mathrm{CO}(1-0)} = 0.77 \pm 0.14$]{boogaard20}. The adoption of a constant excitation correction determined for metal-rich cosmic noon galaxies is  supported by observations of local dwarf galaxies, which show no evidence for systematically different CO excitation conditions compared to more metal-rich local systems \citep{Meier2001, Cormier2014}. CO(1-0) luminosities are subsequently transformed into molecular gas masses using a metallicity-dependent CO-to-H$_2$ conversion factor. As discussed and motivated in detail in \citet{Langan2026}, we adopt the \citet{Accurso2017} CO-to-H$_2$ prescription throughout the ACE paper series, as it provides the best agreement with independent dynamical mass constraints for the ACE sample. All reported molecular gas masses include the contribution from Helium and heavy elements.

\begin{figure*}[ht]
        \centering
        \includegraphics[width=.9\textwidth]{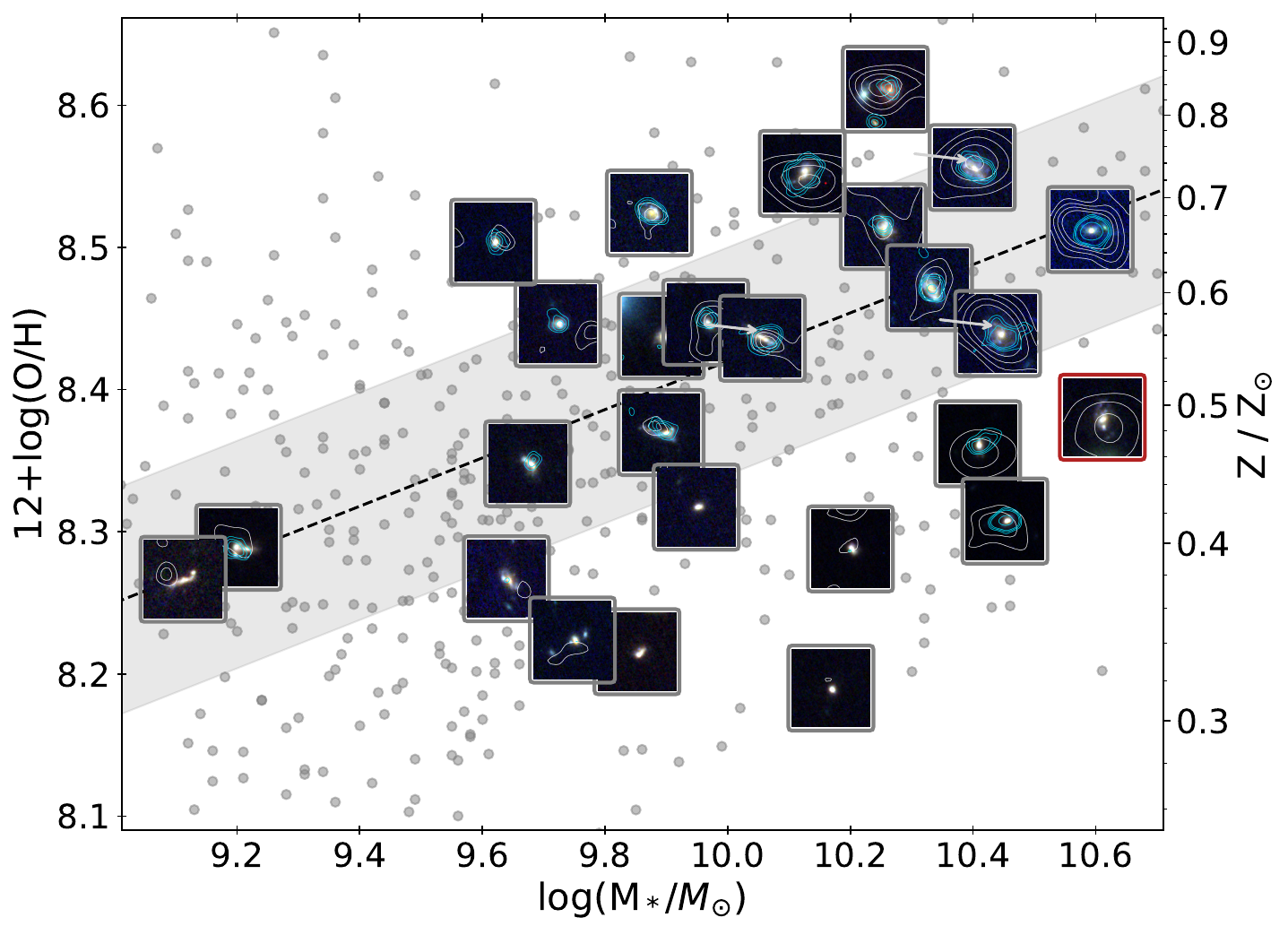} 
    \caption{The mass-metallicity relation of ACE (shown by image cutouts) and the parent sample of MOSDEF (grey points).
    RGBs are NIRCam F150W-F277W-F444W colors. White and cyan contours are CO(3-2) and Band 7 continuum data shown in Figure~\ref{fig:cutouts}. Grey dots in the background are the MOSDEF sample at $z\sim 2$, and the line and shaded region show the mass-metallicity relation at $z\sim 2.3$ \citep{sanders21}, same as in Figure~\ref{fig:MS_MZR}. Three objects are arbitrarily shifted, indicated with a white arrow, as they were overlapping another object. The galaxy with red border is ID 2672, which was not observed as part of ACE and only has CO(3-2) data from \citet{Sanders2023}. The right vertical axis shows metallicities relative to the solar metallicity, assuming $12+\log(\rm{O/H})_{\odot}=8.69$ \citep{asplund09}.}
    \label{fig:MZR_cutouts}
\end{figure*}

\begin{figure*}[ht]
\centering
        \includegraphics[width=0.9\textwidth]{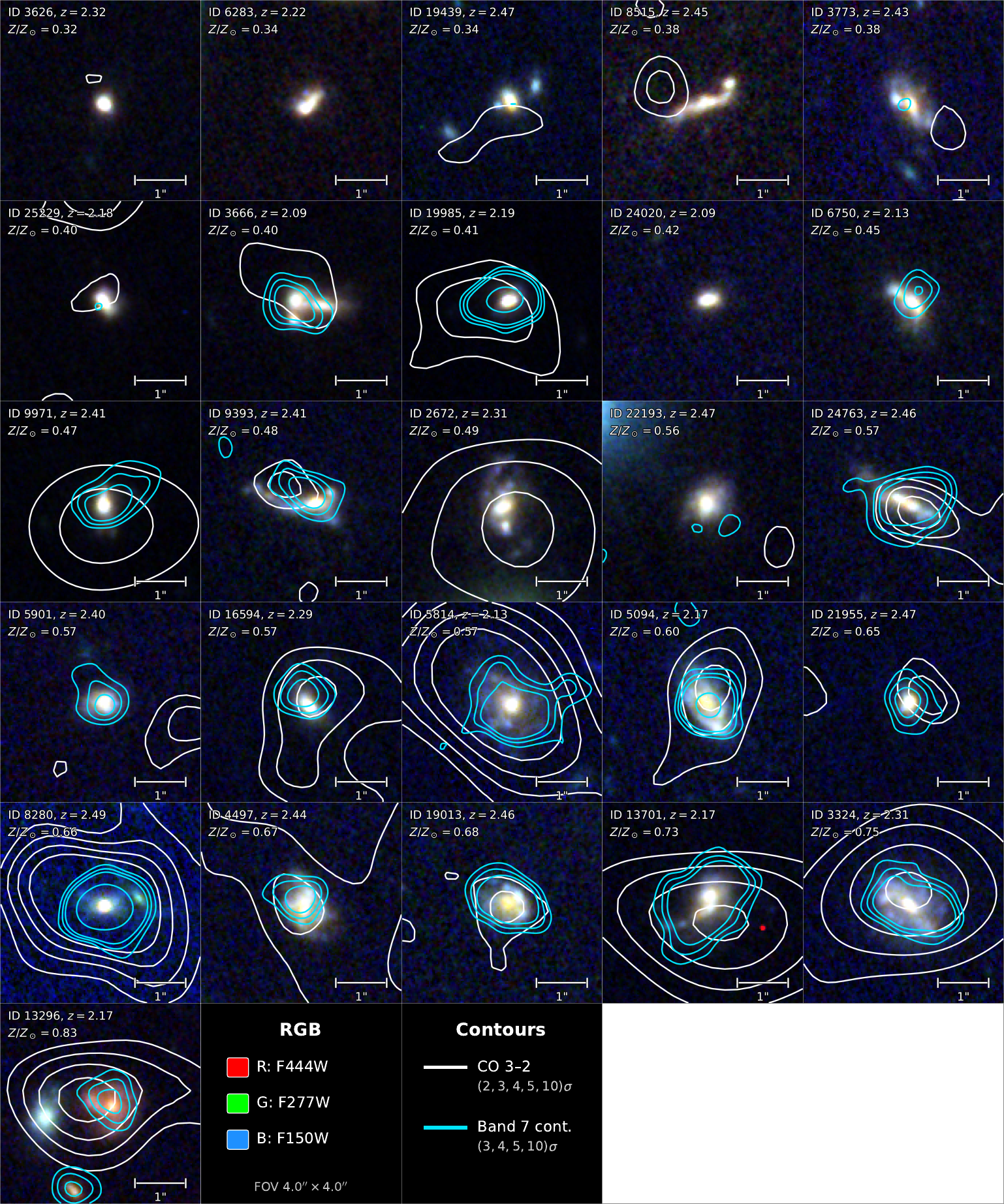}
    \caption{NIRCam F444W/F277W/F150W RGB poststamps of the all the ACE targets with CO(3-2) and Band 7 continuum contours overlaid. The ALMA Band 7 data has a typical beam size of $\sim 0.96''$ \citep{Popescu2026}, whereas the CO(3-2) data has a beam size ranging from 1.5 to $3''$ optimized for each source through different tapering strategies \citep{Langan2026}.
    We note that the integrated spectrum of ID 25229 shows a CO detection (SNR$\sim3$), but this arises from a single peak pixel just above SNR = 3, and as a result, the moment-0 map does not show coherent contours at the corresponding significance level.
    }
    \label{fig:cutouts}
\end{figure*}

\begin{figure*}[ht]
        \centering
        \includegraphics[width=.9\textwidth]{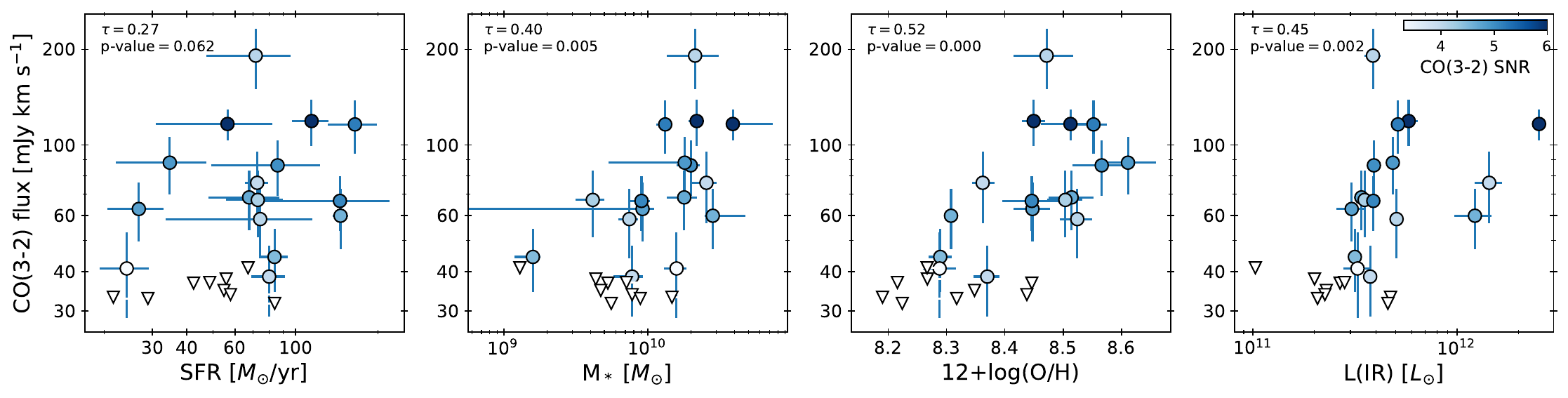} \\
        \includegraphics[width=.9\textwidth]{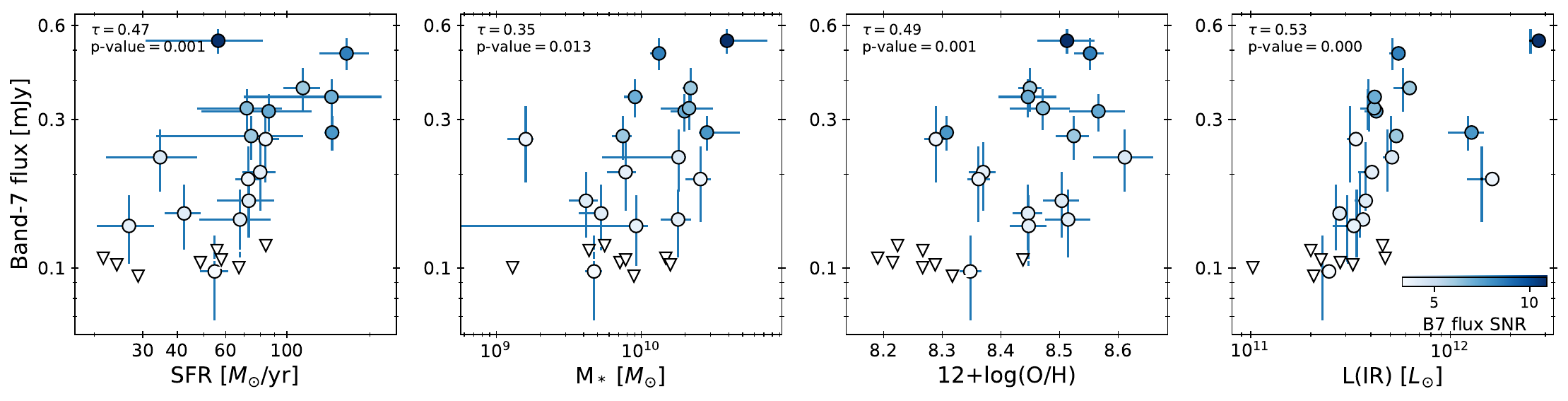}         
    \caption{
    ACE flux measurements of CO(3-2) (top) and Band 7 (bottom) as a function of galaxy properties: (from left to right) dust-corrected H$\alpha$ SFR, stellar mass, metallicity, and IR luminosity. Colors indicate the S/N of the y-axis quantity. Non-detections are shown as $3\sigma$ upper limits with white triangles. Non-parametric correlation coefficient Kendall's tau \citep{Flury_kendall}, accounting for censoring (upper limits), and associated p-values are shown on the top-left corner of each plot.
    The CO detection and flux correlates more strongly with stellar mass and metallicity, compared to SFR and IR luminosity. On the other hand, dust continuum flux in band 7 (and detection rate) is highly correlated with SFR and IR luminosity, while metallicity has no significant correlation with it.
    }\label{fig:CO-B7}
\end{figure*}

\section{Results} \label{sec:results}

\subsection{The lowest metallicity CO and dust detections} \label{sec:lowZ-CO-dust}
\begin{figure*}[t]
\centering
        \includegraphics[width=.9\textwidth]{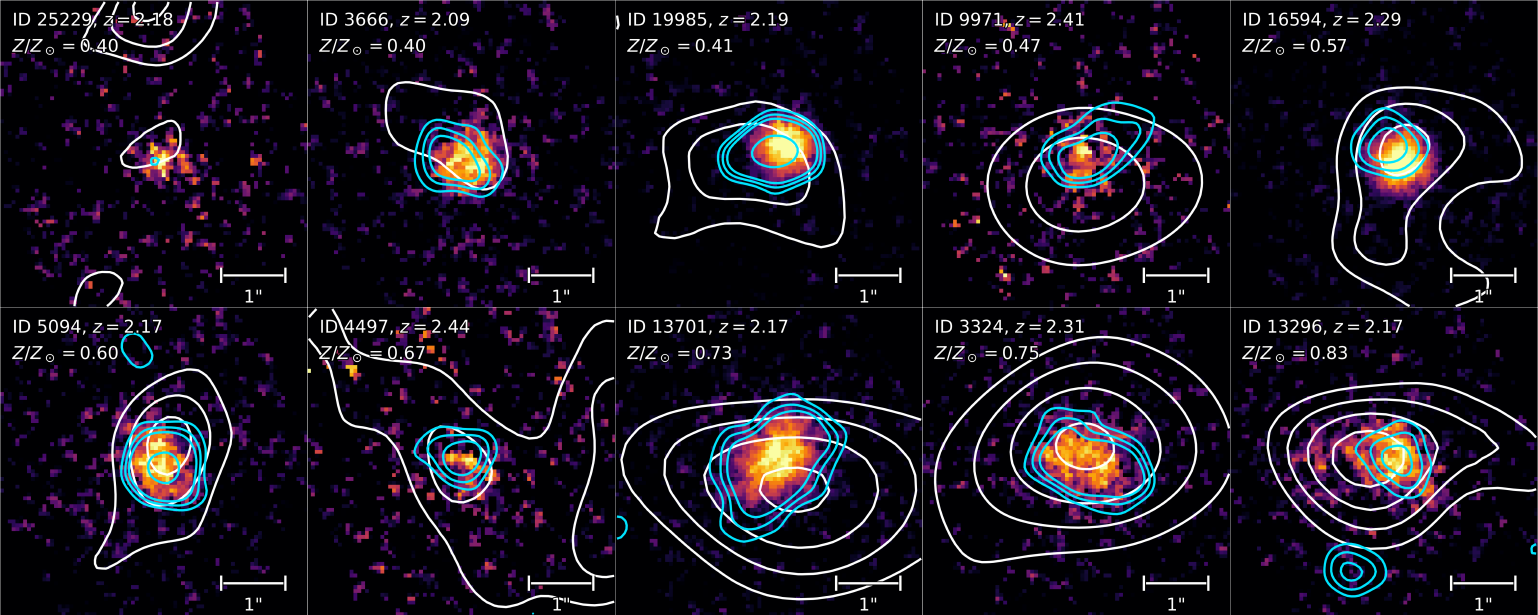} 
    \caption{MIRI 21\,$\mu$m detections, tracing the PAH7.7 complex, in the order of increasing metallicity from top-left to bottom-right. White and cyan contours show CO(3-2) and Band 7 continuum data at 2 (for CO(3-2), 3, 5, 10$\sigma$ levels.}
    \label{fig:miri}
\end{figure*}

ACE detects gas and dust in galaxies at lower metallicities at $z>1$ than ever before (Figure~\ref{fig:MZR_cutouts} and Figure~\ref{fig:cutouts}). Below the metallicity of $\sim 40$\%\,$Z_{\odot}$ there are three CO detections: IDs 25229, 3666, and 19985 that we discuss below in more detail.\footnote{We note that while ID 8515 (lowest mass galaxy in ACE) shows a detection in its Band 3 moment-0 map, we considered it as a non-detection due to the large offset of the peak emission and also lack of Band-7 detection.} All of these three galaxies also have high S/N MIRI 21\,$\mu$m detections, tracing hot dust and PAH emission (Figure~\ref{fig:miri}), setting them as ideal cases for detailed study of dust and gas in low-metallicity galaxies at cosmic noon. Note that all quoted brightnesses, gas and dust fractions and depletion times are presented in detail in \citet{Popescu2026}, \citet{Langan2026}, \citet{Solimano2026}, and \citet{Popping2026} in detail.

Galaxy ID 25229 with $12+\log(\mathrm{O/H})=8.29$ is the lowest metallicity CO detection in the sample with stellar mass of $\log(M_*/M_{\odot})=10.20$ and SFR$_{\rm H\alpha,H\beta}$ of 24\,$M_{\odot}/$yr. It has a molecular gas fraction of $M_{\mathrm{mol}}/M_*\sim 1$ and a gas depletion time of $\log(t_{\mathrm{dep}}/\mathrm{yr})=9$. It is undetected in Band-7, with a low dust mass fraction estimate from SED fitting of $M_{\mathrm{dust}}/M_*\sim 0.001$.

Galaxy ID 3666 is one of the lowest mass and lowest metallicity galaxies in the sample with both CO and Band 7 dust continuum detections with $12+\log(\mathrm{O/H})=8.29$ and $\log(M_*/M_{\odot})=9.20$. This galaxy has a complex morphology, with three components visible in the NIRCam image, which might indicate star-forming clumps or a recent merger. Its molecular gas fraction is very high, $M_{\mathrm{mol}}/M_*\sim 20$, and it has a relatively short depletion time of $\log(t_{\mathrm{dep}}/\mathrm{yr})=8.6$. It is also one of the galaxies with the highest offset from the main-sequence relation, putting it in the starburst region. This galaxy is furthermore the lowest-metallicity dust continuum detection in the sample, with a high dust to mass ratio of $\sim 5-9\%$ depending on whether the dust mass is estimated from the SED fitting (Section~\ref{sec:sed}) or the RJ tail \citep[see][]{Solimano2026}. Such a high dust-to-stellar mass ratio is challenging to reconcile with current models of dust evolution and buildup \citep{Solimano2026}. Given the complex, clumpy morphology of this system (most of the CO(3-2) emission is significantly offset from the three components), geometric effects may contribute to the discrepancy, with the optical emission probing different regions of the galaxy than the dust and molecular gas traced at sub-millimeter wavelengths. Higher-resolution ALMA observations will be essential to determine whether the dust and gas emission is spatially offset from the stellar clumps, thereby clarifying the nature of this source and the origin of its large dust-to-stellar mass ratio.

Galaxy ID 19985 is the third (second) lowest metallicity galaxy with detected CO (detected Band 7) with 42\% solar metallicity. Similar to ID 3666, the centroid of the CO(3-2) emission appears to be significantly spatially offset from the stellar continuum. However, unlike ID 3666, its stellar mass is on the higher end of the ACE sample and has a compact stellar morphology.  This galaxy with the stellar mass of $\log(M_*/M_{\odot})=10.45$, has a molecular mass fraction of $\sim 1.5$, and a dust mass fraction of $0.004-0.005$ depending on the methodology (SED fitting or RJ tail). It is also detected in Herschel PACS and band 6 dust continuum \citep{shivaei22a}. While its metallicity is low, the other parameters of this source is closer to the typical values at these redshifts. Taken together, the dust and gas detections of 19985 and 3666, along with their low metallicities and otherwise distinct galaxy properties, make them ideal case studies for investigating the cold dust and gas properties of low-metallicity galaxies at cosmic noon.

In Figure~\ref{fig:CO-B7}, we further investigate the detection statistics of the ACE sample by presenting the CO(3-2) and Band~7 continuum fluxes as a function of several galaxy properties. The CO(3-2) flux correlates most strongly with metallicity, with a similarly strong trend observed with stellar mass, whereas the correlations with SFR and total IR luminosity are noticeably weaker. We also identify a pronounced decline in the CO detection rate toward lower IR luminosities, with no CO(3-2) detections below $L_{\rm IR}\sim3\times10^{11}\,L_\odot$. In contrast, the Band~7 dust continuum emission displays its strongest correlation with SFR and shows a more gradual decrease in detection rate toward lower IR luminosities than the CO emission.

The majority of the non-detections are concentrated at the low end of the stellar mass, metallicity, SFR, and IR luminosity distributions, suggesting that sensitivity limitations, rather than an intrinsic absence of CO or dust-continuum emission, are the primary driver of the non-detections. This interpretation is supported by the stacking analysis of \citet{Geesink2026}, who detect both CO(3-2) and dust continuum emission in stacks of the lowest-metallicity ACE galaxies (see also \citealt{Solimano2026} and \citealt{Popping2026}).

\subsection{NIRCam and MIRI morphologies of ACE} \label{sec:morphology}

ACE targets have a wide range of morphologies, while selected to be mostly isolated cases. Figure~\ref{fig:MZR_cutouts} shows the NIRCam color cutouts of the galaxies in the MZR plane. While most galaxies are compact, those with the higher mass and metallicity show face-on extended morphologies. The CO and dust continuum detections are in most cases cospatial with the NIRCam morphology (tracing stellar emission). However, there are cases that are peculiar, such as ID 9393. This galaxy with an extended tail in NIRCam images, show an offset dust continuum and CO peak (and even the dust continuum and CO are offset). The offset CO peak can also be seen in other galaxies, including some with higher-S/N detections, such as IDs 5094 and 13701. In cases with low S/N, however, the apparent offset may be due to the limited sensitivity.
There are also cases with a clear displacement between the dust continuum and CO emission, where the dust continuum is associated with a redder NIRCam region, indicating a dustier region, such as IDs 24763 and 13296. The MIRI 21\,$\mu$m peak of ID 13296 is exactly coincident with the peak of the Band-7 continuum (Figure~\ref{fig:miri}), further confirming the dusty nature of this region. Such offsets are expected from clumpy star formation, in which the dust continuum (ALMA) and PAH emission (MIRI) trace the more intensely star-forming and dust-obscured regions, whereas the low-J CO emission better traces the broader gas reservoir. These galaxies are therefore interesting targets for further studies to investigate potentially clumpy star formation, or the more exotic scenario of strong gas outflows.

Among the more massive and metal-rich galaxies in the sample, there are two CO non-detections (IDs 5901 and 22193 with $12+\log(\mathrm{O/H})\sim 8.45$). ID 5901 has a mass of $\log(M_*/M_{\odot})=9.7$ and is detected in dust continuum, while ID 22193 with stellar mass of $\log(M_*/M_{\odot})=9.9$ and IR luminosity of of $\log(L_{IR}/L_{\odot})=11.4$ is the highest-metallicity continuum non-detection in the sample.

We also explored the 21\,$\mu$m MIRI coverage of the ACE sample, as it traces the hot dust and PAH emission, providing interesting additional information on dust species in these galaxies. Out of 12 galaxies that have MIRI F2100W filter coverage, 10 of them have clear detections, shown in Figure~\ref{fig:miri}. Nine of these also have Band-7 cold dust continuum detections ($\sim 270\,\mu$m rest frame). All three lowest-metallicity CO detections discussed in Section~\ref{sec:lowZ-CO-dust} also have MIRI F2100W detections, putting them among the lowest metallicity galaxies with a complete dataset on cold gas and PAHs at high redshifts.
In all cases with Band-7 detections, the emission closely follows the MIRI 21\,$\mu$m emission (rest-frame $\sim 6-7\,\mu$m), showing remarkable agreement between the morphologies of the cold and warm dust continuum (e.g., objects 5094, 13701, 13296, and 3324 in Figure~\ref{fig:miri}). This indicates that even though the ACE observations were not designed to spatially resolve these galaxies, the structure we see in Band-7 is real. 
A followup ALMA+JWST program (ALMA 2025.1.01403.S and JWST 9400; PI: I. Shivaei) will provide high-resolution ALMA Band-7 data and multi-band MIRI images for 6 of the ACE galaxies to perform a detailed morphological study of PAH and cold dust distribution in these galaxies.  

\begin{figure*}[ht]
        \centering
        \includegraphics[width=.9\textwidth]{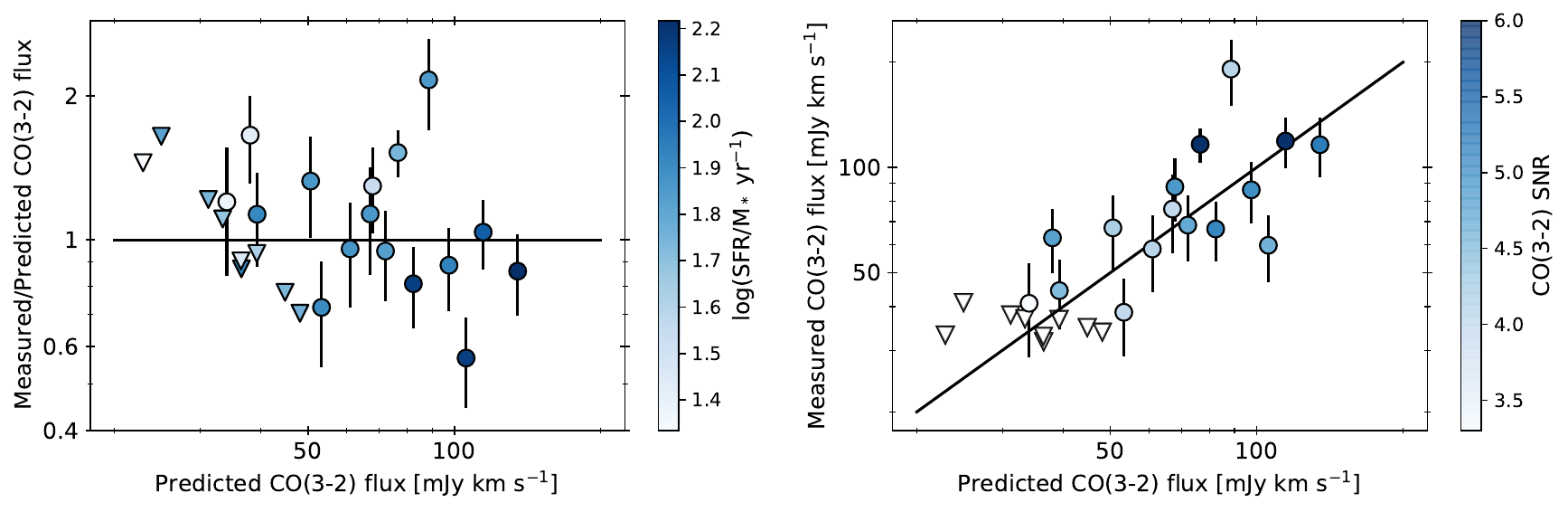} 
    \caption{
    ACE measured CO(3-2) fluxes compared to the predicted values from scaling relations of \cite{tacconi18}, a metallicity-dependent $\alpha_{\mathrm{CO}}$ relation \citep{Bolatto2013}, and excitation correction of $L'_{\rm CO(3-2)}/L'_{\rm CO(1-0)}=0.77$ \citep{boogaard20}. Non-detections are shown with triangles as $3\sigma$ upper limits. Data points are color coded with dust-corrected {\Ha} SFRs on the left and with signal to noise ratio (S/N) of the CO line on the right panel. Among various galaxy properties, the measured-to-predicted CO residuals correlate most strongly with SFR, which is expected owing to the correlation between the excitation of CO and SFR. }\label{fig:CO_predicted_measured}
\end{figure*}

\begin{figure}[ht]
        \centering
        \includegraphics[width=.9\columnwidth]{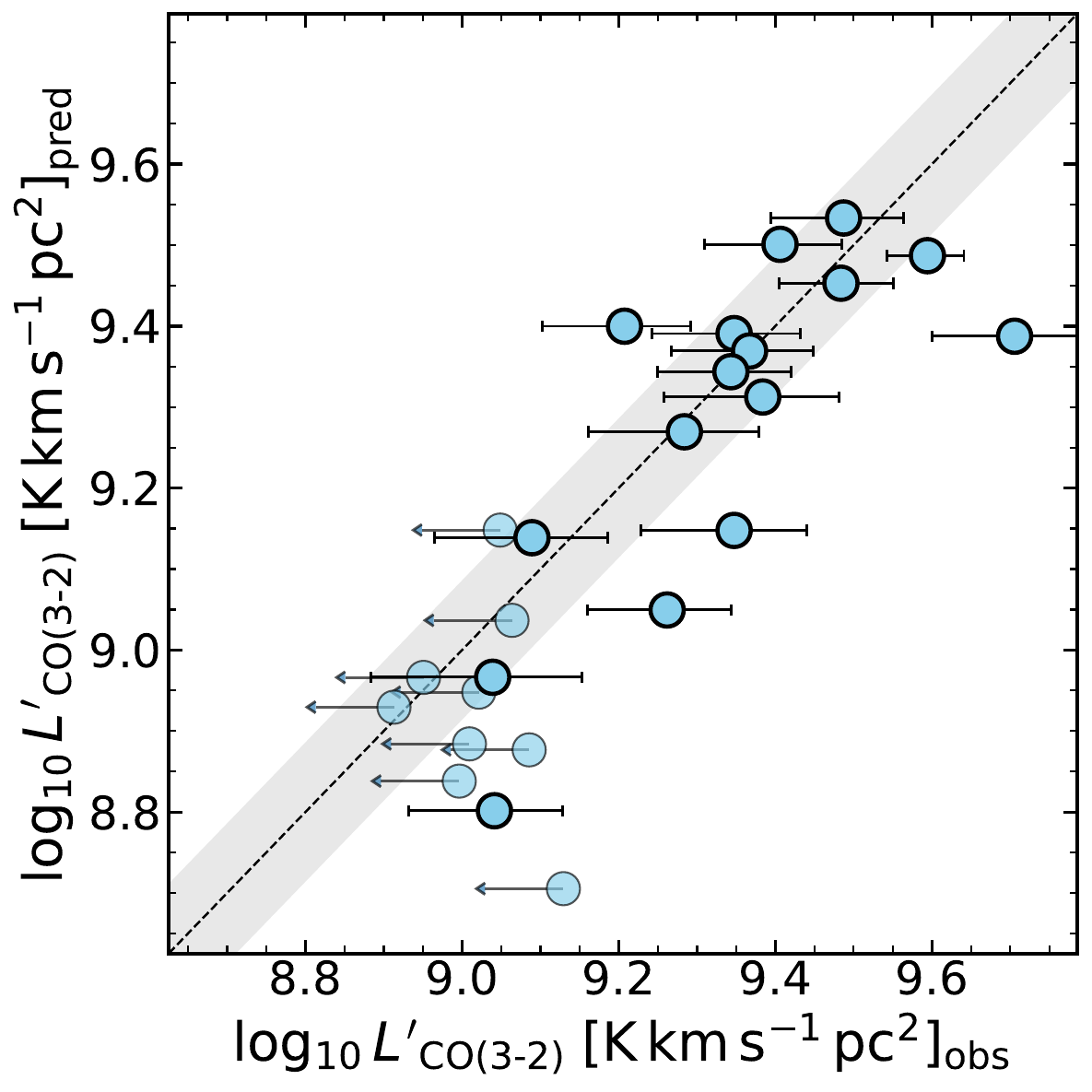} 
    \caption{
    ACE predicted CO(3-2) line-luminosities (y-axis) from Equation~\ref{eq:fitting_function} compared to the observed values (x-axis) , linking the CO(3-2) line luminosity of galaxies at $z\sim 2- 2.5$ to their stellar mass, SFR, and metallicity. 
    }\label{fig:CO_ACE_New_predicted_measured}
\end{figure}

\subsection{An empirical CO(3-2)–SFR–mass–metallicity scaling relation} \label{sec:COpred}
In Figure~\ref{fig:CO_predicted_measured} we compare the measured fluxes of the sample with commonly-used predictions in the literature. We estimate the expected CO(3-2) brightness for our targets using \cite{tacconi18} scaling relation that depends on stellar mass, SFR deviation from the main-sequence \citep{speagle14}, and redshift. Here we use the metallicity dependent $\alpha_{\rm CO}$ relation from \citet{Bolatto2013}\footnote{We note that for the ACE survey we typically use the $\alpha_{\rm CO}$ trend with metallicity and offset from the star-forming main-sequence presented by \citet{Accurso2017}, motivated by dynamical arguments presented in \citet{Langan2026}. We here use the \citet{Bolatto2013} prescription as this was used at the time of proposal writing.} and an excitation correction between CO(3-2) and CO(1-0) line luminosity of 0.77 \citep{boogaard20}. The ratio of measured to predicted flux of CO(3-2) is most strongly correlated with SFR with a Kendall correlation coefficient $\tau$ for censored data \citep{Akritas96,flores04} of $-0.26$ and a p-value of 0.07.  This is not unexpected, since the excitation of CO will vary modestly from source-to-source and is known to correlate with SFR both at cosmic noon and in the local universe \citep[e.g.,][]{boogaard20, keenan2025b, keenan2025a}. The CO flux of galaxies with highest SFR in our sample are over-predicted by the scaling relation by as much as a factor of 2, however, still consistent with the predictions within 3$\sigma$.

The approach presented in Figure~\ref{fig:CO_predicted_measured} relies on assumptions regarding both the CO-to-H$_2$ conversion factor, $\alpha_{\rm CO}$ (which can vary by factors of a few, or more at low metallicity), and, to a lesser extent, the excitation correction between the CO(3-2) and CO($1$--$0$) transitions (varying between 0.5 to 1.0, \citealt{boogaard20}, resulting in only $\sim 30\%$ change in the predicted CO(3-2)). While these assumptions are necessary for deriving molecular gas masses, they introduce additional sources of uncertainty when comparing observed and predicted CO fluxes. To facilitate future studies of galaxies in the low-mass and low-metallicity regime uniquely probed by ACE, it is therefore useful to derive an empirical scaling relation that directly links the observed CO(3-2) luminosity to global galaxy properties. We model the CO(3-2) luminosity as a logarithmic function of stellar mass, SFR, and gas-phase metallicity. The relation is fitted using a Bayesian linear regression framework, where the intercept, slopes, and intrinsic scatter ($\sigma$) are treated as free parameters. The posterior distributions are sampled using a Markov Chain Monte Carlo (MCMC) approach. For galaxies with significant CO detections, we adopt a Gaussian likelihood that accounts for the measurement uncertainties on the CO(3-2) luminosities. Non-detections are incorporated through a censored likelihood function, allowing upper limits to contribute to the fit while ensuring that the underlying luminosities remain physically bounded. The resulting best-fit relation is given by
\begin{equation}
\begin{aligned}
\log_{10} L'_{\mathrm{CO(3-2)}} &= (0.40 \pm 0.11)\,\log_{10}(\mathrm{SFR}/\mathrm{M}_\odot\,\mathrm{yr}^{-1}) \\
&\quad + (0.39 \pm 0.08)\,\log_{10}\!\left(\frac{M_*}{10^{10}\,\mathrm{M}_\odot}\right)\\
&\quad + (0.98 \pm 0.23)\,\left(12 + \log\!\left(\mathrm{O/H}\right) - 8.69\right) \\
&\quad + (8.73 \pm 0.22) 
\end{aligned}
\label{eq:fitting_function}
\end{equation}
with an intrinsic scatter of $\sigma = 0.09 \pm 0.02$ and $L'_{\mathrm{CO}(3\text{-}2)}$ in K km/s pc$^2$. In Figure~\ref{fig:CO_ACE_New_predicted_measured} we show the one-to-one relation between the predicted and observed CO(3-2) line-luminosity of the ACE galaxies when adopting Equation~\ref{eq:fitting_function}.  The typical scatter between the predicted and observed CO(3-2) line-luminosity is $\sim0.1$ dex, though we caution that the intrinsic scatter may be underpredicted at low metallicities due to the increasing fraction of upper limits, even when using a survival analysis.  Regardless, the relation offers a practical benchmark to estimate the molecular gas content directly from the observed CO(3-2) luminosity for galaxies at cosmic noon.

\section{Main Science Goals of ACE} \label{sec:goals}

ACE addresses key questions about the cold gas and dust properties of main-sequence, low-metallicity galaxies at cosmic noon. By targeting galaxies across a wide range of stellar masses and extending down to $\sim0.3\,Z_\odot$, it provides some of the lowest-metallicity detections of CO and dust continuum in unlensed galaxies at $z>1$. A coordinated series of papers presents the first scientific results from the survey, each focusing on a key component of the baryon cycle. Here we briefly summarize these works and refer the reader to the individual papers for detailed analyses.

\subsection{Fundamental gas metallicity relation and cold gas mass-loading factors}

In \citet{Langan2026}, we present the CO(3-2) imaging and flux measurements for the ACE sample and derive molecular gas masses, with particular attention to the choice of the metallicity-dependent CO-to-H$_2$ conversion factor from \citet{Accurso2017}. Using these measurements, we investigate the gas fundamental metallicity relation, linking stellar mass, metallicity, and molecular gas content at cosmic noon, as a physically motivated extension of the traditional FMR. Finally, we place constraints on the cold gas mass-loading factors of the ACE galaxies by comparing the observations to gas-regulator and equilibrium models of galaxy evolution.

\subsection{Dust spectral energy distributions and temperatures}

In \citet{Popescu2026}, we present the imaging and continuum flux measurements obtained as part of ACE. Using a combination of the new ALMA Band~7 observations, archival Band~6 data, Band~3 continuum measurements, and available \textit{Herschel} photometry, we characterize the far-infrared and submillimeter spectral energy distributions of the ACE galaxies through modified blackbody modeling. These measurements provide improved constraints on the dust temperature and emissivity index $\beta$ in the previously unexplored regime of low-metallicity and low-mass galaxies at cosmic noon, enabling an investigation of the physical drivers of dust heating and variations in dust properties.

\subsection{Molecular gas scaling relations}

In \citet{Popping2026}, we investigate how molecular gas mass, molecular gas fraction, and depletion time depend on stellar mass, metallicity, specific SFR, and offset from the star-forming main sequence. ACE extends these scaling relations into a region of parameter space that has remained largely unexplored at cosmic noon, allowing us to test the validity of commonly adopted gas scaling relations at lower stellar masses and metallicities than previously possible.

\subsection{Dust mass scaling relations}

In \citet{Solimano2026}, we derive dust masses for the ACE galaxies and investigate their dependence on stellar mass, SFR, specific SFR, and metallicity. By extending these relations to significantly lower metallicities than previously accessible at $z\sim2$, we place new empirical constraints on the processes governing dust buildup in galaxies and the relative importance of stellar dust production and grain growth in the ISM.

\subsection{Dust-to-gas ratio}

In \citet{Geesink2026}, we combine the dust, gas, and metallicity measurements from ACE to investigate the dust-to-molecular-gas ratio as a function of metallicity. This constitutes the first study of the dust-to-gas ratio in unlensed, sub-solar metallicity galaxies at cosmic noon. The resulting constraints provide insight into the balance between dust production, grain growth, and dust destruction processes, and allow direct comparisons with theoretical models of dust evolution. In addition, \citet{Geesink2026} examines the use of dust continuum emission as a tracer of molecular gas mass and CO luminosity in low-metallicity galaxies \citep[e.g.,][]{scoville16}.

\section{Summary} \label{sec:summary} 

The ALMA Chemical Evolution (ACE) survey is the first ALMA Large Program designed to simultaneously characterize the molecular gas, dust, and metal content of a representative sample of low-metallicity star-forming galaxies at cosmic noon. Combining ALMA CO(3-2) and Band-7 dust-continuum observations with robust gas-phase metallicities from MOSDEF and extensive ancillary photometry from \textit{HST}, \textit{Spitzer}, \textit{Herschel}, and \textit{JWST}, ACE extends studies of the baryon cycle to stellar masses and metallicities well below those probed by previous unlensed surveys.

The ACE observations yield 17 CO(3-2) detections and 17 Band~7 continuum detections, corresponding to detection fractions of 65\% and 68\%, respectively. The detected galaxies span a wide range in stellar mass and metallicity, extending to $\sim40\%$ solar metallicity and including the lowest-metallicity CO and dust detections currently known for unlensed galaxies at cosmic noon. We find that CO detectability correlates most strongly with metallicity and stellar mass, while dust continuum detectability is more closely linked to SFR and IR luminosity. The concentration of non-detections at the low-metallicity, low-mass end of the sample, together with the successful stacking analyses presented in the companion papers, indicates that the current census is primarily limited by sensitivity rather than the absence of molecular gas or dust. Beyond the statistical properties of the sample, ACE has uncovered several particularly interesting systems, including low-metallicity galaxies with substantial molecular gas reservoirs, extreme dust-to-stellar mass ratios, and offsets between the stellar, dust, and molecular gas emission. 

We also derive a new empirical scaling relation linking CO(3-2) luminosity directly to stellar mass, SFR, and gas-phase metallicity, calibrated using a censored Bayesian regression that incorporates both detections and non-detections. The relation reproduces the observed CO(3-2) luminosities with an intrinsic scatter of only $\sim0.1$\,dex, though we caution the latter may be underestimated at low luminosities.  This relation therefore offers a practical benchmark for estimating molecular gas content in future studies of low-mass, low-metallicity galaxies at cosmic noon.

The ACE observations presented here establish the observational foundation for the coordinated series of companion papers summarized in Section~\ref{sec:goals}. Together, the ACE survey provides a comprehensive view of the interplay between gas, dust, metals, and star formation at the peak epoch of galaxy assembly and establishes an unprecedented benchmark dataset for studies of galaxy evolution and chemical enrichment at high redshift.

The ACE survey represents a substantial observational investment, combining nearly 200 hours of ALMA observing time across Bands~3, 6, and 7 to establish the first statistically meaningful census of CO and dust continuum emission in sub-solar metallicity galaxies at cosmic noon. While ACE demonstrates that direct detections of both CO and dust continuum emission are now feasible in this regime, it also highlights the rapidly increasing observational cost of probing toward even lower metallicities and lower-mass systems. A significantly enhanced line sensitivity, such as envisioned by the ALMA2040 concept \citep{Facchini2025}, will be crucial for routinely detecting the cold ISM in the metal-poor galaxy population at cosmic noon and even higher redshifts and for extending direct studies of the baryon cycle to even lower metallicities than currently possible.

\begin{acknowledgements}
IS, IL, and MS acknowledge funding from the European Research Council (ERC) under the European Union’s Horizon 2020 research and innovation programme (DistantDust, Grant agreement No. 101117541) and the Atracc\'{i}on de Talento Grant No. 2022-T1/TIC-20472 of the Comunidad de Madrid, Spain.
LAB acknowledges support from the Dutch Research Council (NWO) under grant VI.Veni.242.055 (\url{https://doi.org/10.61686/LAJVP77714}).
RP acknowledges support for this work provided by the NSF through award SOSPA11-005 from the NRAO.
This paper makes use of the following ALMA data: ADS/JAO.ALMA\#2018.1.01128.S, 2024.1.00534.L. ALMA is a partnership of ESO (representing its member states), NSF (USA) and NINS (Japan), together with NRC (Canada), MOST and ASIAA (Taiwan), and KASI (Republic of Korea), in cooperation with the Republic of Chile. The Joint ALMA Observatory is operated by ESO, AUI/NRAO and NAOJ. We acknowledge assistance and computational support provided by Allegro, the European ALMA Regional Center node in the Netherlands.
\end{acknowledgements}

\bibliographystyle{aa}
\bibliography{bibliography}

\FloatBarrier

\appendix
\section{ACE parameters}
In Table~\ref{tab:ace_sample_props} we present the basic properties of the ACE galaxies. CO and continuum line fluxes are presented in Table~\ref{tab:ace_sample}, whereas derived dust and molecular gas masses are presented in \citet{Solimano2026} and \citet{Langan2026}, respectively.
\onecolumn
\begin{landscape}
\scriptsize
\setlength{\tabcolsep}{4pt}
\centering
%\begin{tabular}{cccccccccc}
\begin{longtable}{cccccccccccc}
\caption{ACE sample properties.}
\\ \hline  
ID & RA & Dec & $z$ & 12+log(O/H) & M$_*$[$\times 10^{9} M_{\odot}$] & SFR({\Ha},{\Hb}) [$M_{\odot}$/yr] & SFR(UV) [$M_{\odot}$/yr] & SFR(IR) [$M_{\odot}$/yr] & L(IR)[$\times 10^{11} L_{\odot}$] & F160W mag & $R_e$ [arcsec]\\
\hline
3324 & 150.148407 & 2.21313357 & 2.31 & $8.57^{+0.05}_{-0.05}$ & $19.89^{+4.06}_{-3.17}$ & $85.88\pm36.81$ & $7.37^{+1.98}_{-2.26}$ & $58.57^{+5.46}_{-0.40}$ & $3.94^{+0.37}_{-0.03}$ & 22.44 & $0.59\pm0.01$ \\
3626 & 150.1047363 & 2.21573091 & 2.32 & $8.19^{+0.03}_{-0.03}$ & $14.73^{+1.54}_{-1.64}$ & $21.58\pm1.86$ & $14.68^{+0.10}_{-0.49}$ & $70.71^{+8.90}_{-4.02}$ & $4.75^{+0.60}_{-0.27}$ & 22.98 & $0.09\pm0.00$ \\
3666 & 150.0775299 & 2.21603322 & 2.09 & $8.29^{+0.02}_{-0.02}$ & $1.59^{+0.40}_{-0.21}$ & $83.68\pm9.83$ & $8.40^{+1.52}_{-1.62}$ & $47.32^{+2.11}_{-1.52}$ & $3.18^{+0.14}_{-0.10}$ & 22.05 & $0.43\pm0.00$ \\
3773 & 150.1981506 & 2.21658921 & 2.43 & $8.27^{+0.03}_{-0.03}$ & $4.36^{+0.78}_{-0.68}$ & $55.75\pm14.88$ & $7.54^{+2.13}_{-2.13}$ & $29.95^{+3.25}_{-0.03}$ & $2.01^{+0.22}_{-0.00}$ & 23.54 & $0.44\pm0.02$ \\
4497 & 150.0714722 & 2.22387266 & 2.44 & $8.52^{+0.04}_{-0.04}$ & $17.92^{+4.35}_{-4.13}$ & $67.61\pm19.56$ & $6.53^{+1.85}_{-3.82}$ & $50.84^{+6.47}_{-2.10}$ & $3.42^{+0.57}_{-0.14}$ & 22.74 & $0.39\pm0.01$ \\
5094 & 150.1403656 & 2.23018551 & 2.17 & $8.47^{+0.05}_{-0.06}$ & $21.32^{+7.71}_{-9.93}$ & $71.46\pm24.27$ & $5.31^{+1.64}_{-0.27}$ & $58.00^{+1.45}_{-11.69}$ & $3.90^{+0.10}_{-0.79}$ & 23.00 & $0.60\pm0.02$ \\
5814 & 150.1691284 & 2.23839569 & 2.13 & $8.45^{+0.02}_{-0.02}$ & $21.88^{+2.66}_{-2.25}$ & $114.30\pm17.38$ & $8.00^{+1.38}_{-0.94}$ & $86.83^{+6.31}_{-11.98}$ & $5.83^{+0.42}_{-0.81}$ & 22.23 & $0.56\pm0.02$ \\
5901 & 150.1894074 & 2.23814201 & 2.40 & $8.45^{+0.03}_{-0.03}$ & $5.29^{+1.58}_{-0.67}$ & $42.39\pm6.34$ & $5.20^{+1.95}_{-1.16}$ & $39.95^{+2.50}_{-3.75}$ & $2.68^{+0.17}_{-0.25}$ & 23.44 & $0.24\pm0.01$ \\
6283 & 150.1152802 & 2.24191332 & 2.22 & $8.22^{+0.03}_{-0.03}$ & $7.11^{+1.65}_{-1.59}$ & $48.53\pm4.78$ & $8.29^{+1.47}_{-0.17}$ & $42.11^{+1.87}_{-8.57}$ & $2.83^{+0.13}_{-0.58}$ & 23.22 & $0.18\pm0.01$ \\
6750 & 150.1630249 & 2.24749112 & 2.13 & $8.35^{+0.02}_{-0.02}$ & $4.72^{+0.62}_{-0.61}$ & $54.71\pm6.43$ & $7.46^{+1.75}_{-1.82}$ & $34.26^{+1.11}_{-0.91}$ & $2.30^{+0.08}_{-0.06}$ & 22.81 & $0.36\pm0.01$ \\
8280 & 150.0865021 & 2.26431680 & 2.49 & $8.51^{+0.05}_{-0.05}$ & $38.98^{+1.97}_{-35.00}$ & $56.35\pm25.60$ & $4.45^{+2.92}_{-22.88}$ & $377.77^{+97.33}_{-98.10}$ & $25.38^{+6.54}_{-6.59}$ & 23.43 & $0.49\pm0.06$ \\
8515 & 150.1844788 & 2.26626229 & 2.45 & $8.27^{+0.02}_{-0.03}$ & $1.30^{+0.58}_{-0.19}$ & $66.99\pm10.81$ & $9.53^{+1.32}_{-1.56}$ & $15.34^{+1.06}_{-1.09}$ & $1.03^{+0.07}_{-0.07}$ & 23.36 & $0.56\pm0.02$ \\
9393 & 150.1635437 & 2.27508616 & 2.41 & $8.37^{+0.02}_{-0.02}$ & $7.79^{+2.03}_{-1.43}$ & $80.01\pm11.13$ & $7.18^{+1.54}_{-2.54}$ & $56.12^{+3.72}_{-4.94}$ & $3.77^{+0.25}_{-0.33}$ & 22.86 & $0.60\pm0.03$ \\
9971 & 150.1435394 & 2.28179717 & 2.41 & $8.36^{+0.02}_{-0.02}$ & $25.56^{+5.33}_{-4.47}$ & $72.34\pm7.18$ & $12.82^{+0.39}_{-0.76}$ & $215.30^{+28.53}_{-37.95}$ & $14.46^{+1.92}_{-2.55}$ & 22.47 & $0.20\pm0.01$ \\
13296 & 150.1150971 & 2.31529069 & 2.17 & $8.61^{+0.05}_{-0.05}$ & $18.07^{+12.70}_{-2.24}$ & $34.68\pm12.59$ & $6.59^{+3.05}_{-1.11}$ & $72.59^{+17.57}_{-24.70}$ & $4.88^{+1.18}_{-1.66}$ & 22.56 & $0.41\pm0.01$ \\
13701 & 150.1127167 & 2.31943941 & 2.17 & $8.55^{+0.02}_{-0.03}$ & $13.23^{+1.80}_{-0.66}$ & $164.61\pm33.70$ & $7.89^{+0.93}_{-1.28}$ & $76.96^{+1.04}_{-4.97}$ & $5.17^{+0.07}_{-0.33}$ & 22.14 & $0.49\pm0.01$ \\
16594 & 150.1249542 & 2.35021901 & 2.29 & $8.45^{+0.03}_{-0.03}$ & $9.21^{+10.69}_{-1.87}$ & $26.73\pm6.25$ & $6.49^{+2.98}_{-4.71}$ & $45.58^{+20.80}_{-6.60}$ & $3.06^{+1.40}_{-0.44}$ & 23.03 & $0.27\pm0.01$ \\
19013 & 150.1119995 & 2.37263298 & 2.46 & $8.52^{+0.03}_{-0.03}$ & $7.45^{+1.22}_{-1.15}$ & $74.18\pm40.60$ & $9.57^{+1.78}_{-1.35}$ & $75.61^{+5.35}_{-6.42}$ & $5.08^{+0.36}_{-0.43}$ & 23.10 & $0.36\pm0.01$ \\
19439 & 150.1015015 & 2.37672329 & 2.47 & $8.22^{+0.02}_{-0.02}$ & $5.58^{+1.11}_{-1.04}$ & $83.84\pm12.75$ & $3.10^{+0.91}_{-1.22}$ & $68.53^{+4.94}_{-1.70}$ & $4.60^{+0.33}_{-0.11}$ & 23.57 & $0.18\pm0.01$ \\
19985 & 150.0603485 & 2.38277268 & 2.19 & $8.31^{+0.01}_{-0.01}$ & $28.31^{+3.50}_{-19.49}$ & $145.84\pm9.19$ & $14.32^{+3.89}_{-2.78}$ & $183.46^{+29.66}_{-105.90}$ & $12.33^{+1.99}_{-7.11}$ & 21.87 & $0.16\pm0.00$ \\
21955 & 150.0953522 & 2.40281034 & 2.47 & $8.50^{+0.03}_{-0.03}$ & $4.15^{+1.00}_{-0.87}$ & $72.71\pm16.99$ & $2.79^{+0.59}_{-0.75}$ & $52.70^{+5.22}_{-0.79}$ & $3.54^{+0.35}_{-0.05}$ & 23.84 & $0.17\pm0.01$ \\
22193 & 150.0854187 & 2.40595770 & 2.47 & $8.44^{+0.04}_{-0.04}$ & $7.79^{+3.33}_{-1.28}$ & $57.81\pm25.66$ & $7.62^{+2.32}_{-1.75}$ & $33.66^{+0.65}_{-16.30}$ & $2.26^{+0.04}_{-1.10}$ & 22.66 & $0.32\pm0.01$ \\
24020 & 150.1151733 & 2.42553139 & 2.09 & $8.32^{+0.03}_{-0.03}$ & $8.88^{+1.53}_{-1.65}$ & $28.86\pm8.89$ & $2.73^{+0.63}_{-0.72}$ & $31.04^{+1.87}_{-0.13}$ & $2.09^{+0.13}_{-0.01}$ & 23.63 & $0.13\pm0.01$ \\
24763 & 150.0567017 & 2.43466282 & 2.46 & $8.45^{+0.05}_{-0.05}$ & $9.05^{+1.45}_{-1.26}$ & $145.09\pm75.50$ & $8.30^{+2.20}_{-0.71}$ & $58.13^{+6.76}_{-7.98}$ & $3.91^{+0.45}_{-0.54}$ & 22.75 & $0.62\pm0.02$ \\
25229 & 150.1084137 & 2.43971491 & 2.18 & $8.29^{+0.03}_{-0.03}$ & $15.85^{+2.84}_{-2.72}$ & $24.15\pm4.94$ & $9.68^{+0.56}_{-1.37}$ & $48.73^{+10.53}_{-4.14}$ & $3.27^{+0.71}_{-0.28}$ & 22.79 & $0.15\pm0.00$ \\
\end{longtable}
\tablefoot{Columns are: ID: 3D-HST v4.0 catalog ID. RA and DEC: coordinates correspond to the optical (3D-HST) coordinates. $z$: MOSFIRE spectroscopic redshift. $12+\log(\mathrm{O/H})$: metallicity and its corresponding error using multiple optical strong nebular emission lines, depending on availability (but at least [O{\sc iii}], {\Ha}, {\Hb}, [N{\sc ii}]) using \cite{sanders26} calibration. M$_*$: stellar mass from \texttt{Prospector}. SFR({\Ha},{\Hb}): Balmer-decrement dust-corrected SFR from {\Ha} \citep{shivaei22a}. SFR(UV): SFR from 1600\,$\AA$ luminosity using \cite{kennicutt12} calibration. SFR(IR): SFR from IR luminosity using \cite{kennicutt12} calibration. L(IR): IR luminosity at 8-1000\,micron from \texttt{Prospector} best-fit SED. F160W mag: AB magnitude in HST F160W filter. $R_e$: effective radius in HST F160W filter from \cite{vanderwel14}.
For CO luminosity and molecular gas mass estimates, we refer the reader to \citet{Langan2026}; for ALMA dust continuum fluxes, we refer the reader to \citet{Popescu2026}; for dust mass estimates, we refer the reader to \citet{Solimano2026}.}
\label{tab:ace_sample_props}
\end{landscape}

\section{Other versions of the main-sequence relation} \label{sec:otherMS}

\begin{figure*}[ht]
        \centering
        \includegraphics[width=.9\textwidth]{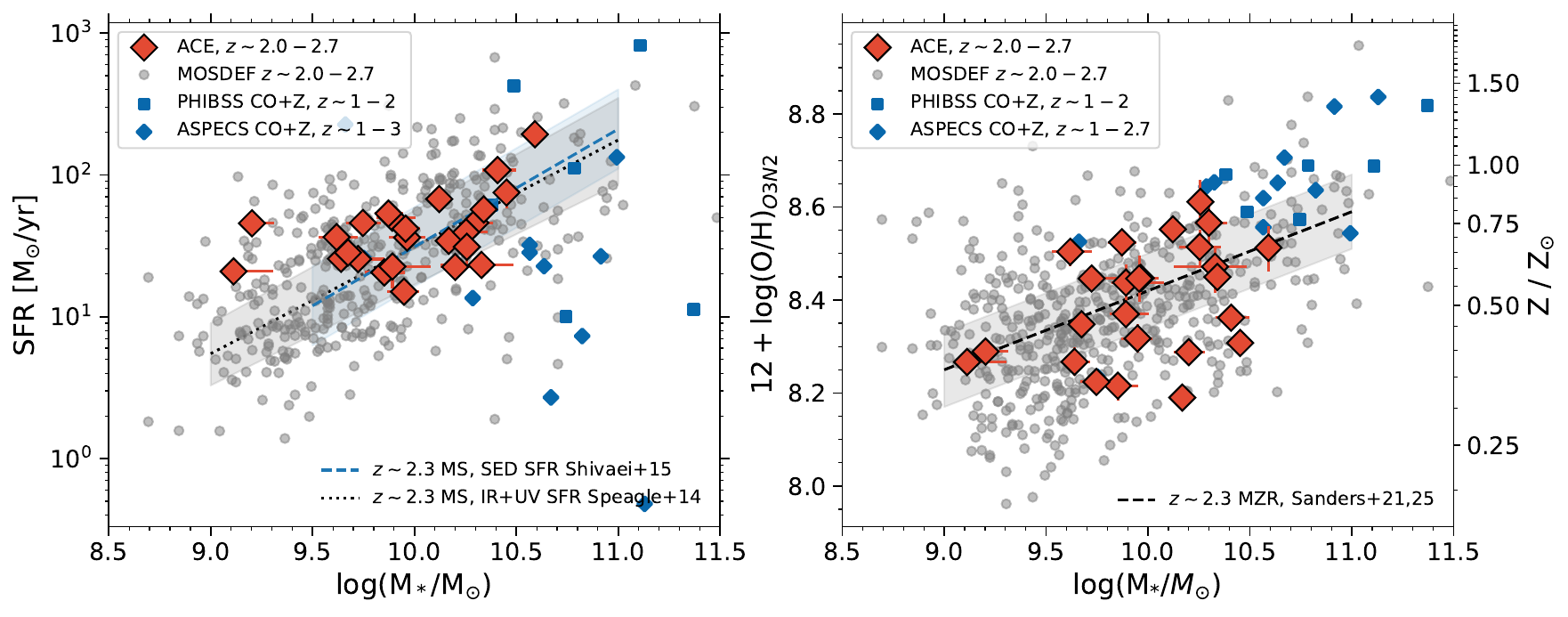} 
    \caption{\label{fig:other-ms}
    Star forming main sequence (left) and mass-metallicity relation (right) for the ACE sample (red diamonds), and its parent sample of the MOSDEF galaxies at $z=2.0-2.7$ (grey circles). Symbols are the same as in Figure~\ref{fig:MS_MZR}, except for the SFRs of ACE. In this figure, SFRs are derived from best-fit SED models (UV to IR, using energy balance). The SED-inferred SFRs compared to either the SED-SFR based main-sequence of \cite{shivaei15b} or the IR+UV SFR based main-sequence of \cite{speagle14} show a much reduced fraction of ACE galaxies above the scatter of the main sequence.
    }
\end{figure*}

As mentioned in Section~\ref{sec:MS}, adopting different SFR indicators/calculations and different mass-SFR main-sequence relations, the ratio of ``starburst'' (i.e., above the main sequence) ACE galaxies varies. 
The {\Ha} main-sequence relation of \cite{shivaei15b} (Figure~\ref{fig:MS_MZR}) indicated 76\% (19 galaxies) above the scatter of the main-sequence relation. On the other hand, using the SFR(IR)+SFR(UV) and the \citet{speagle14} main sequence relation (also based on SFR(IR)+SFR(UV)), 64\% (16 galaxies) are above the main-sequence. Here, in Figure~\ref{fig:other-ms}, we show that assuming the SED-inferred SFRs and the SED-based main-sequence of \citep{shivaei15} (and scatter), only 20\% (5 galaxies) are above the main sequence. 

\end{document}